\documentclass[twocolumn,prd,aps,longbibliography,superscriptaddress,preprintnumbers,tightenlines,showpacs,nofootinbib,amsfonts,amsmath]{revtex4-2}

\pdfoutput=1

\usepackage{epsfig}
\usepackage{graphics}
\usepackage{graphicx}
\usepackage{amsmath,amssymb,mathrsfs}
\usepackage{amsfonts}
\usepackage[usenames,dvipsnames]{xcolor}
\usepackage{xcolor}
\usepackage{wasysym}
\usepackage{times}
\usepackage{newtxmath}
\usepackage{gensymb}
\usepackage{appendix}
\usepackage{listings}
\usepackage{url}
\usepackage[normalem]{ulem}
\usepackage{alltt}
\usepackage[colorlinks=true,
urlcolor=blue]{hyperref}
\usepackage{cleveref}
\usepackage{longtable}
\usepackage{enumitem}
\setlist{nosep}
\usepackage{color}
\usepackage{calc}
\usepackage{tensor}
\usepackage{bm}
\usepackage{times}
\usepackage{multirow}
\usepackage{float}
\usepackage{dcolumn}
\usepackage[nolist,nohyperlinks]{acronym}
\usepackage{xspace}
\usepackage[english]{babel}
\usepackage[abs]{overpic}
\usepackage{pict2e}
\allowdisplaybreaks[1]
\usepackage[utf8]{inputenc}
\usepackage{gensymb}
\usepackage{bm}
\usepackage{stackengine}
\usepackage{boldline,multirow}
\usepackage{braket}
\usepackage{longtable}
\usepackage{orcidlink}

\usepackage{tabularx}
\usepackage{rotating}

\Crefname{figure}{Fig.}{Figs.}

\def\mR{\mathcal{R}}
\def\mP{\mathcal{P}}
\def\mN{\mathcal{N}}
\DeclareMathOperator{\Order}{\mathcal{O}}

\newcommand{\ibar}{{\bar{\imath}}}
\newcommand{\jbar}{{\bar{\jmath}}}

\definecolor{dodgerblue}{HTML}{1E90FF}
\definecolor{viennared}{HTML}{DA0A14}

\hypersetup{citecolor=dodgerblue}

\begin{document}

\title{
	Worldtube excision method for intermediate-mass-ratio inspirals: self-consistent evolution in a scalar-charge model 
}
\newcommand{\cornell}{\affiliation{Cornell Center for Astrophysics and Planetary
		Science, Cornell University, Ithaca, New York 14853, USA}}
\newcommand{\caltech}{\affiliation{Theoretical Astrophysics, Walter Burke
		Institute for Theoretical Physics, California Institute of Technology,
		Pasadena, California 91125, USA}}
\newcommand{\aei}{\affiliation{Max Planck Institute for Gravitational Physics
		(Albert Einstein Institute), Am M{\"u}hlenberg 1, 14476 Potsdam, Germany}}
\newcommand{\fullerton}{\affiliation{Nicholas and Lee Begovich Center for
		Gravitational-Wave Physics and Astronomy, California State University
		Fullerton, Fullerton, California 92831, USA}}
\newcommand{\southampton}{\affiliation{School of Mathematical Sciences and STAG Research Centre, University of Southampton, Southampton, SO17 1BJ, United Kingdom}}
\newcommand{\tata}{\affiliation{International Centre for Theoretical Sciences, Tata Institute of Fundamental Research, Bangalore 560089, India}}
\newcommand{\coimbra}{\affiliation{CFisUC, Department of Physics, University of Coimbra, 3004-516 Coimbra, Portugal}}

\author{Nikolas A. Wittek \orcidlink{0000-0001-8575-5450}} \aei
\author{Adam Pound \orcidlink{0000-0001-9446-0638}} \southampton
\author{Harald P. Pfeiffer \orcidlink{0000-0001-9288-519X}} \aei
\author{Leor Barack \orcidlink{0000-0003-4742-9413}} \southampton
\date{\today}

\begin{abstract}
This is a third installment in a program to develop a method for alleviating the scale disparity in binary black hole simulations with mass ratios in the intermediate astrophysical range, where simulation cost is prohibitive while purely perturbative methods may not be adequate. The method is based on excising a ``worldtube'' around the smaller object, much larger than the object itself, replacing it with an analytical model that approximates a tidally deformed black hole. Previously [Phys.~Rev.~D {\bf 108}, 024041 (2023)] we have tested the idea in a toy model of a scalar charge in a fixed circular geodesic orbit around a Schwarzschild black hole, solving for the massless Klein-Gordon field in 3+1 dimensions on the SpECTRE platform. Here we take the significant further step of allowing the orbit to evolve radiatively, in a self-consistent manner, under the effect of back-reaction from the scalar field. We compute the inspiral orbit and the emitted scalar-field waveform, showing a good agreement with perturbative calculations in the adiabatic approximation. We also demonstrate how our simulations accurately resolve post-adiabatic effects (for which we do not have perturbative results).  In this work we focus on quasi-circular inspirals. Our implementation will shortly be publicly accessible in the SpECTRE numerical relativity code. 

\end{abstract}

\maketitle
\acrodef{NR}{numerical relativity}
\acrodef{GW}{gravitational wave}
\acrodef{BBH}{binary black hole}
\acrodef{BH}{black hole}

\newcommand{\NR}[0]{\ac{NR}\xspace}
\newcommand{\BBH}[0]{\ac{BBH}\xspace}
\newcommand{\BH}[0]{\ac{BH}\xspace}

\newcommand{\citeme}[0]{{\color{purple}{Citation!}}}

\section{Introduction} \label{sec:Intro}

Inspiralling binary black holes (BBHs) will remain prime targets for gravitational-wave searches as we approach the era of third-generation instruments and LISA (the Laser Interferometer Space Antenna). Precision modelling of BBH signals over the full parameter space of expected sources remains a high priority task  \cite{LISAConsortiumWaveformWorkingGroup:2023arg}. Unique difficulties are posed in the intermediate mass-ratios regime, where Numerical Relativity (NR) simulations become less efficient while perturbative methods may not be adequate. This work continues the program initiated in Dhesi {\it et al.}~\cite{Dhesi:2021yje} (henceforth {\it Paper I}) and Wittek {\it et al.}~\cite{Wittek:2023nyi} ({\it Paper II}) aimed at developing a synergistic approach to the problem, combining NR techniques with methods in black hole perturbation theory. The general idea is to alleviate the scale disparity that hampers NR simulations by excising a large region around the smaller black hole (BH), inside which an approximate analytical solution is used, representing a tidally perturbed BH geometry. The smallest lengthscale on the numerical domain is now that of the excised sphere (a ``worldtube'' in spacetime), rather than the scale of the smaller body. As a result, the Courant-Friedrich-Lewy (CFL) stability limit on the timestep of the numerical simulation is relaxed, with a commensurate gain in computational efficiency. 

Paper I laid out the basic framework and tested it in a simple scalar-field model in 1+1 dimensions. In this toy model, reviewed further below, the smaller BH is replaced with a point particle endowed with scalar charge, which sources a (massless) scalar field, assumed to satisfy the Klein-Gordon equation on the fixed geometry of the large object, taken to be a Schwarzschild black hole. The scalar charge in Paper I was taken to move on a fixed circular geodesic orbit around the BH, with both gravitational and scalar-field back-reaction forces ignored.  Paper I was focused on exploring various techniques for matching the numerical field outside the excision worldtube to the analytically prescribed solution inside it. It also investigated and quantified the scaling of the model error with the worldtube size, using two independent numerical implementation schemes. 

Paper II applied the worldtube idea in full 3+1 dimensions, still working with a scalar-field toy model and a fixed circular geodesic source.  The problem was reformulated as an initial-boundary evolution problem suitable for implementation on the SpECTRE platform \cite{spectrecode}, and a completely new implementation code was developed. The paper detailed the construction of a suitable approximate analytical solution inside the worldtube, and devised a procedure for fixing remaining, {\it a priori} unknown degrees of freedom using dynamical matching to the external numerical solution across the worldtube's boundary. The convergence of the numerical solutions with worldtube size was quantified and shown to agree with theoretical expectations. Detailed comparisons were made with analytical solutions in limiting cases, and with numerical results from other simulations, showing a reassuring agreement. 

In the current work we make a crucial step towards the physical BBH problem by relaxing the condition that the scalar charge is moving on a fixed geodesic orbit, and instead allowing the orbit to evolve radiatively, solving the sourced field equation in a self-consistent manner.  This requires substantial adaptations in both formulation and code infrastructure. The analytical model inside the worldtube must be generalised to allow for the source's acceleration as it moves in its inspiral trajectory around the large BH. The architecture of the numerical domain must be significantly modified, too. In particular, our evolution code employs a discontinuous  Galerkin (DG) scheme with several hundred DG elements that are deformed to fit the domain structure using a series of smooth coordinate maps, and these must now become time-dependent. 

We begin in Sec.~\ref{sec:review} with a general summary of the worldtube method. Section \ref{sec:Maps} details our numerical method, including the construction of time-dependent coordinate maps for generic orbits, and the procedure for matching numerical data to the analytical solution across the worldtube's boundary. In Sec.~\ref{sec:Puncture} we give a generalized approximate analytical model for the field inside the worldtube, allowing for source acceleration. Section \ref{sec:Evolution} describes in detail the procedure employed to perform a self-consistent evolution of the sourced field equations, coupled to the particle's equation of motion. Since, at each timestep, the analytical field inside the worldtube depends on the particle's acceleration, which itself is determined from the field that we are attempting to calculate, the acceleration equations take an implicit form. We describe an iterative scheme developed to deal with this problem.
 
Section \ref{sec:Results} contains a sample of illustrative results from our numerical simulations. We show examples of inspiral orbits and emitted scalar-field waveforms, tracking the evolution all through the inspiral, plunge and ringdown phases. We use invariant diagnostics--the adiabaticity parameter and total orbital phase---to perform quantitative tests against accurate perturbative calculations in the adiabatic approximation, showing excellent agreement. We show, furthermore, how our simulations resolve post-adiabatic information.   We explore in detail the scaling of numerical error with worldtube size and with the number of iterations of the acceleration equation, in both cases confirming the expected convergence.  Section \ref{sec:Conclusions} summarizes our results and discusses forthcoming steps in our program. 

To the best of our knowledge, our work is the first to report a fully self-consistent evolution in the scalar-field model. Previously, Diener {\it et al.} \cite{Diener:2011cc} have studied the radiative evolution of orbits in the same model, using an alternative method---the so called ``effective source'' approach (whose relation to our worldtube method is discussed in Sec.~III.C of Paper I). However, in that work it was found that, in order to achieve a numerically stable evolution, certain terms (involving time derivatives of the acceleration) had to be ignored in the equations that couple the particle's equation of motion to the local analytical approximation. For that reason, we were unable to perform a detailed comparison to our results. 

The rest of this introduction reviews the scalar-field toy model employed in this work. Throughout the paper we use geometrized units, with $G=c=1$. We use Latin indices to denote spatial tensor components and Greek indices for spacetime components.

\subsection{Scalar-field toy model}

We consider a Schwarzschild BH of mass $M$ orbited by a pointlike particle carrying a scalar charge $q$ and mass $\mu\ll M$. The particle sources a (test) scalar field $\Psi$, assumed to be governed by the massless Klein-Gordon equation
\begin{equation}\label{eq: KG with source}
	g^{\mu \nu} \nabla_\mu \nabla_\nu \Psi = -4 \pi q \int \frac{\delta^4 ( x^{\alpha} - x^{\alpha}_p (\tau))}{\sqrt{-g}} d\tau,
\end{equation}
and subject to the usual retarded boundary conditions at null infinity and on the event horizon. 
In Eq.~(\ref{eq: KG with source}), $g^{\mu \nu}$ is the inverse Schwarzschild metric and $\nabla_\mu$ is the covariant derivative compatible with it. $x^{\alpha}_p(\tau)$ describes the particle's worldline, parameterized in terms of proper time $\tau$. The worldline itself satisfies the equation of motion
\begin{equation}\label{eq:cov_evolution}
	u^\beta \nabla_\beta ( \mu u_\alpha) = q \nabla_\alpha \Psi^\mR,
\end{equation}
where $u^\alpha\! := \!\frac{d x_p^\alpha}{d\tau}$ is the tangent four-velocity, and $\Psi^\mR$ is the Detweiler-Whiting regular piece of $\Psi$ (`R field') at the position of the particle. On the left-hand side here is the covariant derivative of the particle's four-momentum along the orbit, and the right-hand side represents the back-reaction force from the particle's own scalar field, known as self-force. Equations (\ref{eq: KG with source}) and (\ref{eq:cov_evolution}), together with a prescription for constructing $\Psi^\mR$ out of $\Psi$, form a closed coupled set of ``self-consistent'' evolution equations, whose solution we aim to obtain. This solution is uniquely determined once initial conditions are given in the form of $x^\alpha$ and $u^\alpha$ at an initial time, together with initial data for $\Psi$.

It is useful to split Eq.\ (\ref{eq:cov_evolution}) into its components orthogonal and tangent to $u^\alpha$, respectively given by 
\begin{align}\label{eq:cov_evolution_split}
	u^\beta \nabla_\beta (u_\alpha) &= \frac{q}{\mu} (\delta^\beta_\alpha+u^\beta u_\alpha)\nabla_\beta \Psi^\mR,\\
    \frac{d\mu}{d\tau} &= -q u^\alpha \nabla_\alpha \Psi^\mR.
\end{align}
The first equation describes the self-acceleration of the scalar charge on the Schwarzschild background due to the scalar-field back reaction. The second equation can be immediately integrated to yield 
\begin{equation}\label{eq:dynamic_mass}
	\mu = \mu_0 - q\Psi^\mR,
\end{equation}
which describes the evolution of the particle's mass over time due to exchange of energy with the ambient scalar field.

From Eq.\ (\ref{eq:cov_evolution_split}) and the fact that $\nabla_\beta\Psi^\mR\propto q/M^2$, we see that the magnitude of the self-acceleration is controlled by the dimensionless parameter
\begin{equation}\label{eq:eps}
\epsilon: = \frac{q^2}{\mu_0 M},
\end{equation}
which plays the role of the (small) mass ratio in the analogous BBH problem. 
We assume $\epsilon\ll 1$, in order to ensure that the orbital evolution is slow during the inspiral, as in the BBH case. In practice, $\mu$ changes by a few percent at most during the systems studied here, so the distinction between $\mu$ and $\mu_0$ in Eq.~(\ref{eq:eps}) is subdominant.
In this work we also completely neglect the {\em gravitational} back-reaction on the particle's motion.

\section{Summary of Worldtube Method} \label{sec:review}

In Paper II we developed a technique for solving the field equation (\ref{eq: KG with source}) with a source corresponding to a scalar charge on a fixed, circular geodesic orbit. Much of the infrastructure of Paper I carries over to our present work, so we start with a summary of that infrastructure. %

We describe the trajectory of the scalar charge using $x^i_p(t)$ in Kerr-Schild (KS) coordinates {$t, x^i$} associated with the BH. For the Schwarzschild black hole considered here, the horizon is at $r=2M$ where the radius in KS coordinates is given by
  \begin{equation}
    r=\left(\delta_{ij}x^ix^j\right)^{1/2}.
  \end{equation}
A KS coordinate sphere, centered on $x_p^i(t)$, is excised from the computational domain. We refer to the spacetime boundary of the excised region as {\em the worldtube}, denoted by $\Gamma$. By construction, the scalar charge is always at the center of the spherical excision sphere. Outside the worldtube, we solve the homogeneous Klein-Gordon equation
\begin{equation}\label{eq:KG_vacuum}
	g^{\mu \nu} \nabla_\mu \nabla_\nu \Psi^{\mN} = 0, 
\end{equation}
with 3+1 dimensional numerical relativity methods. The superscript $\mN$ denotes this numerical solution (as distinguished from the fields $\Psi^\mP$ and $\Psi^\mR$ defined below). To facilitate numerical implementation, Eq.~\eqref{eq:KG_vacuum} is reduced to first order in space and time by introducing the following auxiliary variables \cite{Scheel:2003vs}:
\begin{subequations}\label{eq: evolution reduction}
	\begin{align}
		\Pi &= - \alpha^{-1} (\partial_t \Psi^{\mN} - \beta^i \partial_i \Psi^{\mN}), \\
		\Phi_i &= \partial_i \Psi^{\mN},
	\end{align}
\end{subequations}
where $\alpha$ and $\beta^i$ are, respectively, the lapse function and shift vector of the background metric. The coupled evolution equations for $\Pi$ and $\Phi_i$ are given in Eqs.~(8) of Paper II. They are solved using SpECTRE \cite{spectrecode} in 3+1 dimensions using a nodal discontinuous Galerkin (DG) scheme.

In the vicinity of the charge, an approximate particular solution to the inhomogeneous equation~\eqref{eq: KG with source} is given by the puncture field $\Psi^\mP$. It is constructed as an approximation to the Detweiler-Whiting singular field \cite{Detweiler:2002mi} and expressed as a power series in coordinate distance from $x^i_p(t)$. In Paper II, we derived $\Psi^\mP$ for circular geodesic orbits; here, in Section ~\ref{sec:Puncture}, we extend it to generic, accelerated equatorial orbits.

The residual field $\Psi^\mR = \Psi - \Psi^\mP$ approximately solves the homogeneous Klein-Gordon equation in the worldtube's interior. Out perturbative approximation of the interior solution consists of expanding $\Psi^\mR$ and its time derivative as a Taylor Series truncated at order $n$. For $n=1$, these read
\begin{align}\label{eq:psi_expansions}
	\Psi^\mR(t, x^i) &= \Psi^\mR_0(t) + \Psi^\mR_i(t) \rho n^i + \Order (\rho^2), \\
  \label{eq:partialtPsi}
	\partial_t \Psi^\mR(t, x^i) &= (\partial_t\Psi^\mR)_0(t) + (\partial_t\Psi^\mR)_i(t) \rho n^i + \Order (\rho^2).
\end{align}
Different to Paper II, we write this expansion in {\it inertial} KS coordinates $x^i$. We define the displacement to the particle as $\Delta x^{i} := x^i - x^i_p$, the KS spatial distance by $\rho := \sqrt{\delta_{i j} \Delta x^{i} \Delta x^{j}}$ and normal vector through $n^i := \Delta x^i / \rho$. The boundary of the worldtube is located at $\rho=R$ (for some constant $R$). Because Eq.~(\ref{eq:partialtPsi}) expands the inertial time-derivative around a time-dependent expansion point $x_p^i(t)$, the coefficients on the right-hand side of Eq.~(\ref{eq:partialtPsi}) are not the time-derivatives of the coefficients in Eq.~(\ref{eq:psi_expansions}), i.e. $d\Psi_0^\mR/dt \neq (\partial_t\Psi^\mR)_0$.

The essence of the worldtube scheme lies in determining the unknown expansion coefficients in Eqs.~\eqref{eq:psi_expansions} and~\eqref{eq:partialtPsi} dynamically during the evolution. Most of the coefficients are determined from a continuity condition at the worldtube's boundary $\Gamma$, which matches the exterior solution $\Psi^\mN$ to the interior, residual solution $\Psi^\mR$ at each time step:
\begin{align}\label{eq:continuity_condition}
	\Psi^\mR  &\overset{\Gamma}{=} \Psi^\mN - \Psi^\mP, \\  \label{eq:continuity_condition2}
	\partial_t \Psi^\mR  &\overset{\Gamma}{=} \partial_t \Psi^\mN - \partial _t \Psi^\mP.
  \end{align}
This matching is done mode by mode in a multipole expansion, using a procedure described around Eq.~(33) of Paper II. For expansion orders $n > 1$, to fully determine all coefficients one must additionally use further constraints coming from the requirement that $\Psi^\mR$ solves the vacuum Klein-Gordon equations. As described in Paper II, one arrives at ODEs in time, to be solved along with the evolution equation. 

Once fully determined, the expansions~\eqref{eq:psi_expansions} and~(\ref{eq:partialtPsi}) are used to provide boundary conditions to the DG evolution at the worldtube boundary. The exact conditions are derived in Section IV C of Paper II and remain unchanged in this work.

The errors of various quantities in the simulation are expected to scale with the worldtube radius $R$ according to a power law.  Paper II derives the following predictions:
\begin{align}
	\text{Error in } \Psi^\mathcal{N}(x^i):&\quad \mathcal{O}(R^{n+2}), \label{eq:predicted_error_psiN}\\
	\text{Error in } \Psi^\mathcal{R}(x^i_p):&\quad \mathcal{O}(R^{n+1}),\label{eq:predicted_error_psiR}\\
	\text{Error in } \partial_\alpha\Psi^\mathcal{R}(x^i_p):&\quad \mathcal{O}(R^{n}).\label{eq:predicted_error_dpsiR}
\end{align}
The validity of these scaling relations was illustrated numerically in Paper II for a particle on a fixed, circular geodesic orbit with radius $r_0 = 5 M$. 

In the next three sections we describe the extension of the above scheme to radiatively evolving orbits. This involves (i) the addition of time-dependent maps to the code, able to track the particle on generic orbits; (ii) the generalization of the puncture field to generic orbits; and (iii) the derivation of an iterative scheme to accommodate the new puncture field. We restrict ourselves to the first-order expansion case, $n=1$.

\section{Time-Dependent Maps for Generic Orbits}\label{sec:Maps}

\subsection{Coordinate Frames}
\begin{figure*}
	\includegraphics[width=\columnwidth]{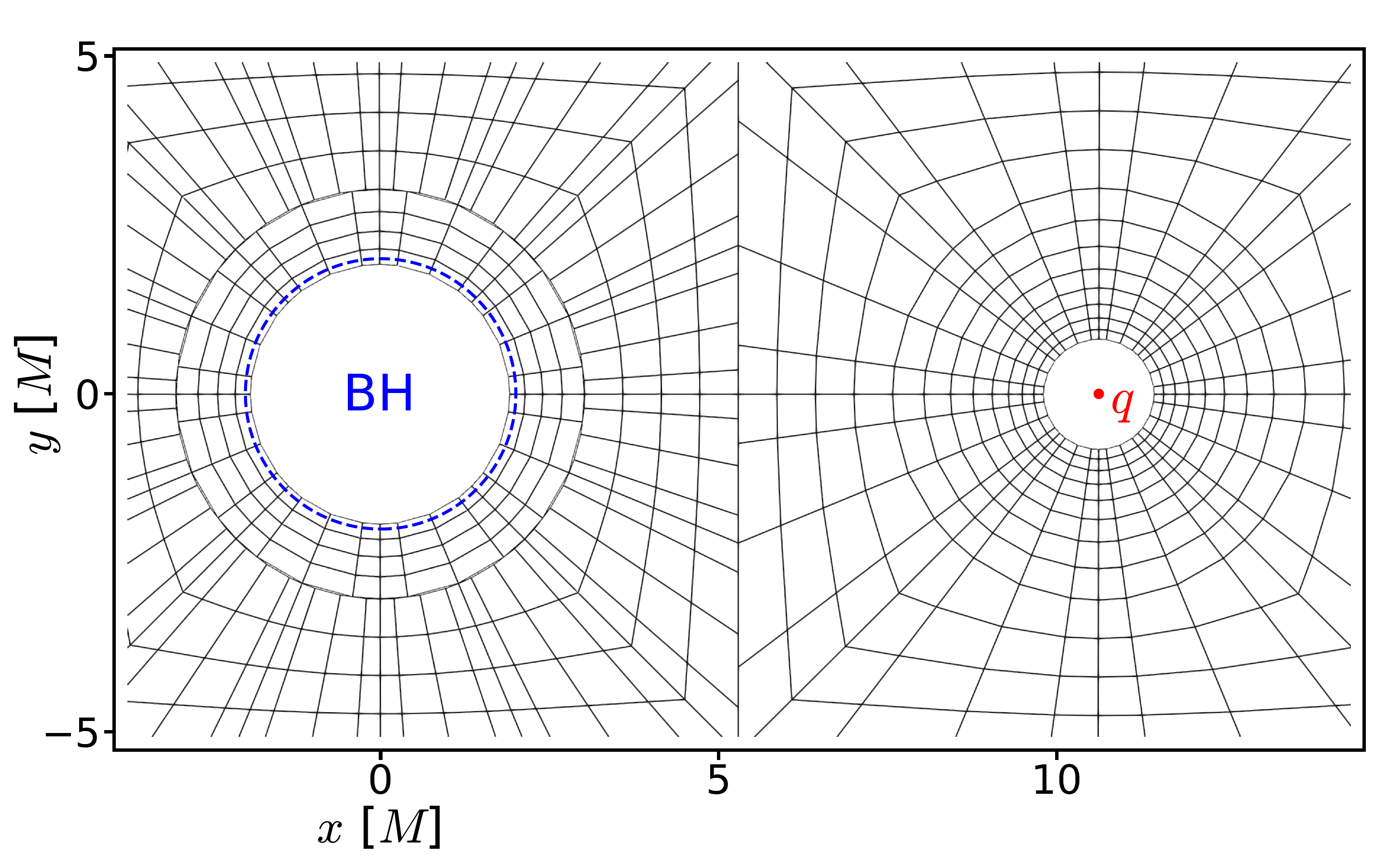}
	\hspace{0.3cm}
	\includegraphics[width=\columnwidth]{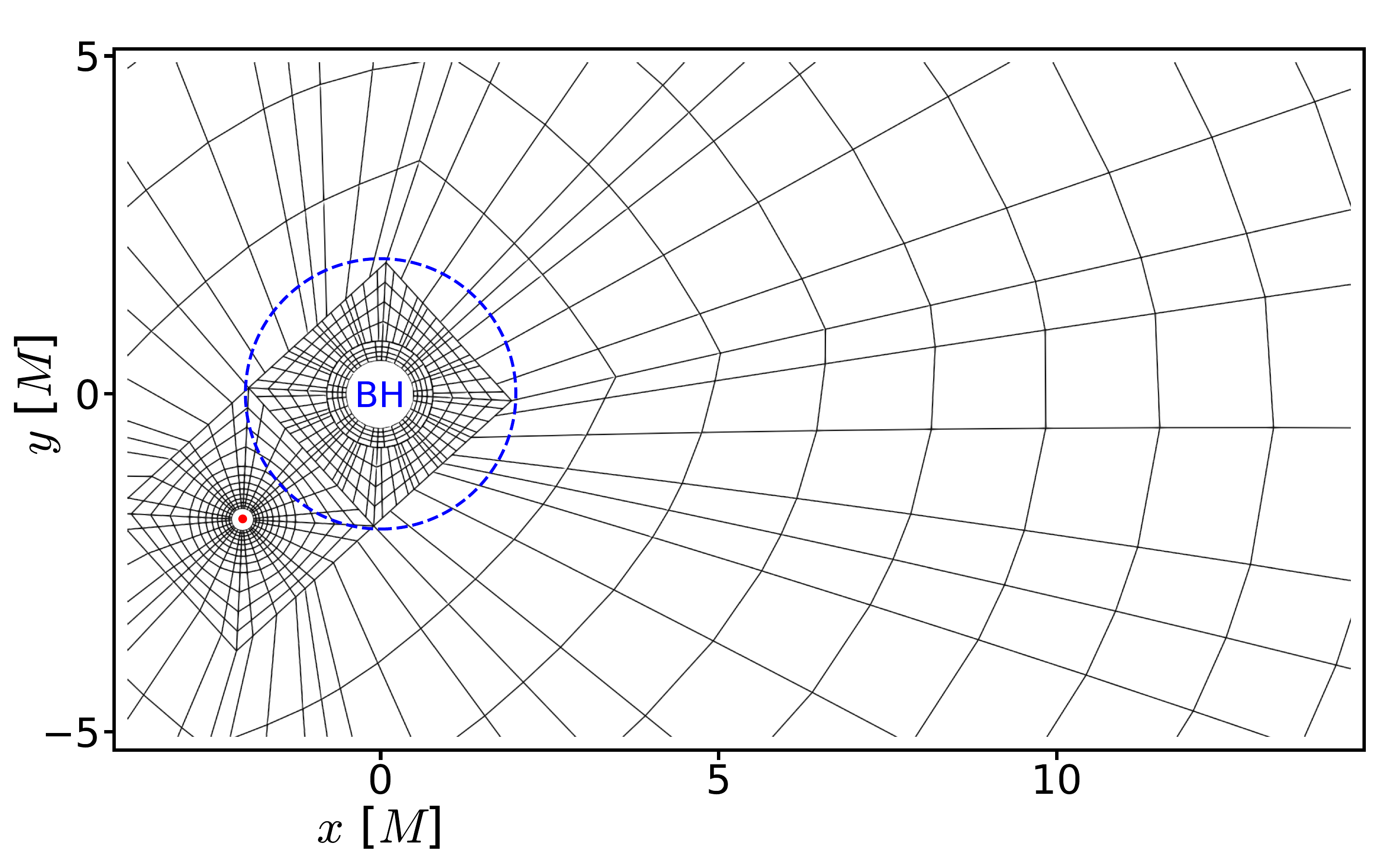}
	\caption{{\it Left: }The DG grid in
    the time-independent grid frame, equivalent to the inertial Kerr-Schild coordinates at the start of a simulation. On the right
    side is the worldtube excision sphere with the scalar charge $q$
    at its center indicated by a red dot. On the left side is the
    excision sphere around the central black hole. The blue ring
    corresponds to the event horizon at $r = 2M$.  The KS coordinates are centered on
    the black hole, and during the evolution the grid rotates around
    this center. {\it Right: }The DG grid in the inertial frame at a
    later time of a simulation, at the same scale as on the left. The
    worldtube excision sphere at the bottom left is close to crossing
    the event horizon at $r = 2M$. A series of time-dependent
    functions map the collocation points from the grid frame as
    depicted in the left panel to the inertial frame by rotating and
    compressing the grid. The approximate value for the phase is $\phi \approx \frac{5 \pi}{4} \pi$, for the orbital radius it is $r_p \approx 2.7 M$, for the worldtube radius $R \approx 0.15 M$, and the black hole excision radius is $\approx 0.5 M$.}
	\label{fig:grid}
\end{figure*}
The computational domain is constructed by combining several hundred DG elements, each containing up to several thousand collocation points. These grid points correspond to the nodal representation of a tensor product of Legendre polynomials using Gauss-Lobatto quadrature. 

The elements are deformed from unit cubes to fit the domain structure using a series of maps. An initial set of time-independent maps transforms them to the so-called grid frame which is co-moving with the grid points. It is depicted in the left panel of Figure~\ref{fig:grid}. We denote the corresponding grid coordinates with a bar,  $x^\ibar$.

A set of time-dependent maps then transform the grid coordinates to the inertial KS coordinates $x^i$ introduced earlier.  These time-dependent maps cause the grid points to move across the spacetime background in the inertial frame, to follow the motion of the scalar point charge. The setup is motivated by BBH evolutions where control systems continually adjust time-dependent parameters in these maps to track the motion and shape of the black holes' apparent horizons~\cite{Scheel:2006gg,Hemberger:2012jz,Scheel:2014ina}.

In this work, we integrate the particle's orbit along with the DG evolution and determine the time-dependent parameters in the maps by demanding that the worldtube is centered on the scalar charge at each time step. This corresponds to the particle physically moving across the KS background with its position fixed at the excision sphere's center. Because $x^i_p(t)$ is determined directly from the ODE Eq.~\eqref{eq:cov_evolution_split}, this setup does not utilize control systems.

In Paper II, we fixed the particle's orbit to be circular. The map from grid to inertial coordinates then amounts to a global rotation with constant angular velocity. We now generalize to a series of time-dependent maps to accommodate generic, equatorial orbits with dynamically adjustable excision radii:

A rotation map controls the angular position of the particle and is applied globally to each DG element according to
\begin{subequations}
	\label{eq:rotation_map}
	\begin{align}
		x &= \bar{x} \cos\phi(t) - \bar{y} \sin\phi(t), \\
		y &= \bar{x} \sin\phi(t) + \bar{y} \cos\phi(t), \\
		z &= \bar{z},
	\end{align}
\end{subequations}
where $\phi(t)$ is the time-dependent rotation angle. The orbital velocity $\dot\phi(t)$ is no longer constant but tracks the particle's orbit. The rotation is always around the z-axis as we fix the particle's orbit in the $xy$-plane.

A compression map stretches grid points according to a time-dependent factor $\lambda(t)$ about a center $C^\ibar$. We define the coordinate distance from $C^\ibar$ in the grid frame to be $\bar r = \sqrt{\delta_{\ibar \jbar}(x^\ibar-C^\ibar)(x^\jbar-C^\jbar)}$. The compression factor falls off linearly in the radial interval $[r_{\mathrm{min}}, r_{\mathrm{max}}]$, and the compression map is given in the piecewise form as
\begin{equation}\label{eq:compression_map}
	x^i = 
	\begin{cases}
		x^\ibar - \lambda(t) \frac{x^\ibar - C^\ibar}{r_{\mathrm{min}}}, & \bar r < r_{\mathrm{min}}, \\[0.5em]
		x^\ibar - \lambda(t) \frac{r_{\mathrm{max}} - \bar r}{r_{\mathrm{max}} - r_{\mathrm{min}}}\frac{x^\ibar - C^\ibar}{\bar r}, & r_{\mathrm{min}} \leq \bar r \leq r_{\mathrm{max}}, \\[0.5em]
		x^\ibar, & \bar r > r_{\mathrm{max}}. 
	\end{cases}
\end{equation}
The Jacobian of this map is discontinuous at $r_{\mathrm{min}}$ and $r_{\mathrm{max}}$. The DG method can handle this as long as these radii are placed at element boundaries. We apply three compression maps to the domain as follows:

A global compression map is centered on the central black hole with the inner radius placed at the so-called envelope $r_{\mathrm{min}} = r_{\mathrm{env}}$ which is chosen several times larger than the initial separation of the two excision spheres. The outer radius $r_{\mathrm{max}}$ is placed at the outer boundary of the domain. The radial separation between the worldtube and the black hole in the inertial frame can then be controlled by adjusting the corresponding parameter $\lambda_r(t)$ which  linearly scales the entire inner portion of the grid. The outer boundary of the domain does not change as the compression factor drops to zero at the outer boundary.

Two additional compression maps are centered on the black hole and the worldtube, respectively, with $r_{\mathrm{min}}$ set to the initial excision sphere radii and $r_{\mathrm{max}}$ placed at the spherical element boundaries surrounding them. We denote the corresponding functions of time as $\lambda_{\mathrm{bh}}(t)$ and $\lambda_{\mathrm{wt}}(t)$, respectively. As the compression map is spherically symmetric, the excision regions remain spherical in the inertial frame.

The combination of all four time-dependent maps allows for separate control of the angular and radial position of the worldtube through $\phi(t)$ and $\lambda_r(t)$, as well as the excision sphere radii through $\lambda_{\mathrm{bh}}(t)$ and $\lambda_{\mathrm{wt}}(t)$. An example of this concatenation of maps is shown in Figure~\ref{fig:grid}. The left figure corresponds to the DG elements in the time-independent grid frame which coincides with the inertial frame the beginning of the simulation. The right figure shows the same grid points transformed to the inertial frame at a later time of the simulation. 

At each time $t$, the DG elements need to be supplied with the value and derivative of the time-dependent parameters $\phi(t)$, $\lambda_r(t)$, $\lambda_{\rm bh}(t)$ and $\lambda_{\rm wt}(t)$ to evaluate the evolution equations at their collocation points in the inertial frame $x^i(t, x^\ibar)$.  At these positions, we compute the metric quantities appearing in the evolution equations. The velocity is needed to transform the time derivative of the evolution equations into the co-moving grid frame as described e.g. in \cite{Scheel:2006gg}. We now show how the values of these time-dependent parameters are determined from the orbit of the scalar charge. 

\subsection{Particle's Position}
At the start of the simulation, all functions of time are set to zero, $\phi=\lambda_r=\lambda_{\rm bh}=\lambda_{\rm wt}=0$, so that grid coordinates coincide with inertial coordinates, $x^i(x^\ibar, t = 0) = x^\ibar$. The worldtube is initially located on the positive $x$-axis with center at an orbital radius $r_0$. 

In Section~\ref{sec:Evolution}, we will derive an ordinary differential equation (ODE) governing the particle's motion. At each time step, we integrate the ODE to calculate the new position $x^i_p(t) = (x_p(t), y_p(t), z_p(t))$ and velocity $\dot{x}^i_p(t)= (\dot{x}_p(t), \dot{y}_p(t), \dot{z}_p(t))$ of the particle in Kerr-Schild coordinates. The time-dependent parameters are then adjusted so that the function from grid to inertial coordinates $x^i(t, x^\ibar)$ maps the center of the worldtube to the current position of the particle,
\begin{equation}\label{eq:mapping_condition}
	x^i(t, x_p^\ibar) = x^i_p(t).
\end{equation}

This condition is satisfied by choosing the following values:
\begin{align}\label{eq:fix_fots}
	\phi(t) &= \arctan\left(y_p(t), x_p(t)\right), \\
	\dot{\phi}(t) &= \frac{x_p(t) \dot{y}_p(t) - y_p(t) \dot{x}_p(t)}{r_p^2(t)}, \\
	\lambda_r(t) &= r_{\mathrm{env}} \left( 1 - \frac{r_p(t)}{r_0}\right), \\
	\dot{\lambda}_r(t) &= \frac{-r_{\mathrm{env}} \dot{r}_p(t)}{r_0},
\end{align}
where we defined the orbital radius of the particle as $r_p(t) = \sqrt{\delta_{ij}x^i_p(t) x^j_p(t)}$, with radial velocity $\dot{r}_p(t) = \delta_{ij} \dot{x}^i_p(t) x^j_p(t) / r_p(t)$.

\subsection{Radii of excision Spheres}\label{sec:Maps/Excision}
The time-dependent map parameters $\lambda_{\mathrm{bh}}(t)$ and $\lambda_{\mathrm{wt}}(t)$ merely modify the size of the excision regions around the centre of the black hole or the scalar charge, respectively, and can be chosen independently of Eq.~\eqref{eq:mapping_condition}.

Our choice for $\lambda_{\rm wt}$ is motivated by observing that the worldtube scheme is more accurate at larger $r_p$, since the expansion terms of the puncture field converge more quickly there. If the orbital radius decreases, the truncation error of the puncture field, and hence that of the regular field too, grow. We expect the error $\varepsilon$ due to the worldtube to scale with $r_p$ as \cite{Dhesi:2021yje}
\begin{equation}
	\varepsilon \sim r_p^{-3(n+1)/2}.
\end{equation}
Recall that $n$ is the expansion order of the scheme, fixed to $n=1$ in this work. The error in the field and its derivatives also scale with the worldtube radius $R$, according to the relations~\eqref{eq:predicted_error_psiN}--\eqref{eq:predicted_error_dpsiR}: $\sim R^{n+1}$ for the field and $\sim R^{n}$ for its derivatives. %
We can keep the error roughly constant as the orbit evolves, by adjusting the worldtube radius $R$ as a function of the changing orbital radius $r_p$. To achieve this we use the power-law relation
\begin{equation}\label{eq:shrink}
	R(t)  = R_0 \left( \frac{r_p(t)}{r_0} \right)^\beta,
\end{equation}
where $R_0$ is the initial excision radius and the exponent $\beta$ can be chosen freely. A value of $\beta = 3/2$ should ensure that the error in $\Psi^\mR$ remains constant; a value of $\beta = 3$ is required to keep the error in the derivatives $\partial_i \Psi^\mR$ constant. For the simulations presented in this work we choose $\beta = 3/2$, as the larger worldtube reduces computational cost.

The excision sphere within the central black hole is assigned an initial radius of $R_0$ = 1.99 M. It is then shrunk using Eq.~\eqref{eq:shrink} with $\beta = 1$. The dynamic shrinking of both excision spheres allows the worldtube to approach and ultimately to pass through the black hole horizon with the grid remaining well-behaved; see the right side of Figure~\ref{fig:grid} for the configuration shortly before the particle passes through the horizon.

Care has to be taken in determining the actual functions of time $\lambda_{\mathrm{bh}}(t)$ and $\lambda_{\mathrm{wt}}(t)$ to match the desired excision sphere radii $R(t)$, as the global compression map governed by $\lambda_r(t)$ has already affected the radii. The appropriate choice to attain an excision sphere radius of $R_{\mathrm{bh/wt}}(t)$ is
\begin{align}\label{eq:excision_fots}
	\lambda_{\mathrm{bh/wt}}(t) &= R_0 + \frac{R_{\mathrm{bh/wt}}(t)r_{\mathrm{env}}}{\lambda_r(t) - r_{\mathrm{env}}}, \\ 
	\dot{\lambda}_{\mathrm{bh/wt}}(t) &= \frac{r_{\mathrm{env}}}{\lambda_r(t) - r_{\mathrm{env}}}\left( \dot{R}_{\mathrm{bh/wt}}(t) + \frac{R_{\mathrm{bh/wt}}(t) \dot{\lambda}_r(t)}{r_{\mathrm{env}} - \lambda_r(t)} \right),
\end{align}
where $R_0 = R_{\mathrm{bh/wt}}(r_0)$ is the excision sphere radius at the start of the simulation.

\section{Puncture Field}\label{sec:Puncture}

Local expansions of the Detweiler-Whiting singular field for a scalar charge are well developed~\cite{Heffernan:2012su, Haas:2006ne, Wardell:2011gb}, as reviewed in Paper II. These have primarily focused on the case of a charge moving on a geodesic, but Refs.~\cite{Pound:2014xva,Heffernan:2017cad}, for example, considered the case of an accelerated source particle. 

Here we start from the results of Ref.~\cite{Pound:2014xva}. That reference provided punctures for gravitational perturbations $h_{\alpha\beta}$ produced by an accelerated point mass $\mu$, but we can readily extract the puncture for our scalar field by noting that the trace of the linear metric perturbation, $h:=g^{\alpha\beta}h_{\alpha\beta}$,  satisfies the same Klein-Gordon equation~\eqref{eq: KG with source} as the scalar field, $g^{\mu\nu}\nabla_\mu\nabla_\nu h = -16\pi \mu \int \frac{\delta^4(x^\alpha-x^\alpha_p(\tau))}{\sqrt{-g}}d\tau$, with the replacement $q\leftrightarrow4\mu$. Therefore we have $\Psi^\mP = \frac{q}{4\mu}h^\mP$.

The resulting puncture takes the form 
\begin{equation}
\Psi^\mP = \Psi^\mP_{\rm geo} + \Psi^\mP_{\rm acc},
\end{equation}
where $\Psi^\mP_{\rm geo}$ is the puncture for a particle on a geodesic, and $\Psi^\mP_{\rm acc}$ is the correction due to the particle's acceleration. The first term, which appeared already in Paper II, is given by
\begin{align}\label{eq:PsiP geo covariant}
\Psi^\mP_{\rm geo} &=  \frac{q}{\lambda s}+ \frac{q\lambda}{6 s^3}(\varrho^2-s^2)  C_{u \sigma u \sigma} +\mathcal{O}(\lambda^2).
\end{align}
Here we have introduced a number of auxiliary quantities. $\lambda:=1$ is used to count powers of distance to the particle. $s$, $\varrho$, and $C_{u\sigma u \sigma}$ are defined from Synge's world function $\sigma(x,\tilde{x})$ and its derivative $\tilde \sigma_{\alpha}:=\tilde \nabla_{\!\alpha}\sigma(x,\tilde{x})$~\cite{J.L.Synge:1960zz}, where we use a tilde to label quantities evaluated on the particle at time $t$, as in $\tilde x^\alpha:=(t,x^i_p(t))$. $\sigma(x,\tilde{x})$ is equal to half the squared geodesic distance between $x^\alpha$ and $\tilde{x}^\alpha$, and its gradient $\tilde \sigma_{\alpha}$ is a directed measure of distance from $\tilde x^\alpha$ to $x^\alpha$. In terms of these, we have defined
\begin{align}
\varrho &:=\tilde \sigma_{\alpha} u^{\alpha},\label{eq:rbar}\\
s&:=\sqrt{(\tilde g^{\alpha \beta}+ u^{\alpha} u^{\beta})\tilde \sigma_\alpha \tilde \sigma_{\beta}}\,,\label{eq:s2}\\
C_{u \sigma u \sigma} &:= \tilde C_{\alpha \beta \mu \nu}  u^{\alpha}\tilde \sigma^{\beta} u^{\mu} \tilde\sigma^{\nu},
\end{align}
where $\tilde C_{\alpha\beta\mu\nu}$ is the Weyl tensor at $\tilde x^\alpha$. Written in an analogous form, the correction to Eq.~\eqref{eq:PsiP geo covariant} due to acceleration reads 
\begin{align}\label{eq:PsiP acc covariant}
\Psi^\mP_{\mathrm{acc}} &= \frac{\lambda^0 f^{\alpha} \sigma_{\alpha} \left(s^2-\varrho^2\right)}{2 s^3} +\lambda\biggl\{\frac{f^{\alpha} \sigma_{\alpha} \left(s^2-\varrho^2\right)}{2 s^3}\nonumber\\
&\quad -\frac{\varrho D_uf^{\alpha} \sigma_{\alpha} \left(\varrho^2-3 s^2\right)}{6 s^3}-\frac{f_{\alpha}f^{\alpha} \left(\varrho^2+s^2\right)}{s}\biggr\}+\mathcal{O}(\lambda^2),
\end{align}
where $f^\alpha$ is the self-force per unit mass, given by the right-hand side of Eq.~\eqref{eq:cov_evolution_split}, and $D_u f^\alpha:=u^{\beta}\nabla_{\beta}f^{\alpha}$ is its covariant derivative along the worldline. Explicitly,
\begin{align}\label{eq:self_force}
	f^\alpha &= \frac{q}{\mu}\left(\tilde g^{\alpha \beta} + u^\alpha u^\beta\right) \partial_\beta \left.\Psi^\mR\right|_{x^i_p}, \\
	u^\beta \nabla_\beta f^\alpha &= \frac{q}{\mu} \left(\dot{f}^\alpha u^0 + \tilde \Gamma^\alpha_{\beta \gamma} u^\beta f^\gamma \right).
\end{align}
In all expressions, it is understood that the four-velocity $u^\alpha$ and self-force per unit mass $f^\alpha$ are evaluated on the worldline at time $t$. We note that the acceleration terms $\Psi^\mP_{\mathrm{acc}}$ depend on the self-force per unit mass $f^\alpha$ at $\Order(\lambda^0)$ and also start to depend on the derivatives $\partial_\beta f^\alpha$ at $\Order(\lambda^1)$.

Starting from the above covariant expansions, we re-expand all quantities in powers of the Kerr-Schild coordinate distance from the particle, $\Delta x^i:=x^i-x^i_p(t)$. That expansion is reviewed in Paper II. Although we use all terms through order $\lambda$ in our numerics, here for brevity we only present the order-$\lambda^{-1}$ and -$\lambda^0$ terms. Our results for those terms are the following:
\begin{widetext}
  \begin{align}\label{eq:PsiP_geo}
	\Psi^\mP_{\rm geo} = \frac{q}{\lambda s_0} + \frac{\lambda^0 qM}{2r_p^8 s_0^3} &\biggl\{ r_p^3 x_p^i \Delta x_p^i x_p^k x_p^l \Delta x^m \Delta x^n \left(3 \delta_{km} \delta_{ln} - 2 \delta_{kl} \delta_{mn}\right) + \left( u^0 \right)^2 \left( r_p^3 \dot{x}_p^i \Delta x^i + 2M r_p x_p^i \Delta x^i  + 2M \dot{x}_p^i x_p^i x_p^j \Delta x^j \right)  \nonumber\\
	 \times &\left[2 r_p x_p^a x_p^b \Delta x^k \Delta x^l \left(2 \delta_{ak} \delta_{bl}  - \delta_{ab} \delta_{kl}\right) + \dot{x}_p^a \Delta x_p^a x_p^k x_p^l \Delta x^m \Delta x^n \left(3 \delta_{km} \delta_{ln} - 2 \delta_{kl} \delta_{mn}\right)\right]\biggr\}+\mathcal{O}(\lambda)
\end{align}
for the geodesic piece and
\begin{align}\label{eq:PsiP_acc}
	\Psi^\mP_{\mathrm{acc}} =&  \frac{\lambda^0 q}{2r_p^9 s_0^3}\Bigg\{ \left( u^0\right)^2\Big[ 2M r_p x^i_p \Delta x^i_p + 2M x^i \dot{x}^i_p x^j_p \Delta x^j  + r_p^3 \dot{x}^i_p \Delta x^i \Big] ^2 \Big[ 2M f^t r_p x^i_p \Delta x^i + f^i \left(r_p^3 \Delta x^i + 2M x^i_p x^j_p \Delta x^j \right) \Big]\nonumber \\
  &\qquad\qquad\quad - 2M f^t r_p^7 s_0^2 x_p^i \Delta x^i - r_p^6 s_0^2 f^i \left(r_p^3 \Delta x^i + 2M x^i_p x^j_p \Delta x^j\right) \Bigg\}+\mathcal{O}(\lambda)
\end{align}
for the correction due to acceleration. Here we have introduced the convention that two repeated upper indices are summed over with a Kronecker delta, i.e. $x^i y^i := x^i y^j \delta_{ij}$. We have also introduced $u^0:=dt/d\tau$, given by
\begin{equation}
	\left( u^0 \right)^2 = \frac{r_p^3}{ r_p^3\left( \dot{x}_p^i \dot{x}_p^i -1 \right) -2M  r_p^{2}   - 4Mr_p x_p^i \dot{x}_p^i- 2M \left(x_p^i \dot{x}_p^i \right)^2},
\end{equation}
and the leading-order term in the coordinate expansion of Eq.~\eqref{eq:s2}, given by
\begin{equation}
	s_0^2 = \Delta x^i \Delta x^i + \frac{2M \left(\Delta x^i x_p^i \right)^2}{r_p^3} + \frac{\left( u^0 \right)^2 \left(r_p^3 \Delta x^i \dot{x}_p^i + 2Mr_p \Delta x^i x_p^i + 2M \Delta x^i x_p^i  \dot{x}_p^j x_p^j \right)^2}{r_p^6 }.
\end{equation}
\newpage
\end{widetext}

\section{Self Consistent Evolution}\label{sec:Evolution}
The motion of a scalar charge subject to the scalar self-force is governed by Eq.~\eqref{eq:cov_evolution_split}. In coordinate form, the spatial components are given by
\begin{equation}\label{eq:coord_evolution}
	(u^0)^2 \ddot{x}^i_p  = \frac{q}{\mu}(g^{i \alpha} - \dot{x}^i_p g^{0 \alpha} ) \partial_\alpha \Psi^\mR - (\Gamma^i_{\beta \gamma} - \dot{x}^i_p \Gamma^0_{\beta \gamma} ) u^\beta u^\gamma,
\end{equation}
where $\Gamma^\alpha_{\beta \gamma}$ are the Christoffel symbols of the second kind and $u^i  = \dot{x}^i_p u^0$. The first term on the right-hand side represents the covariant acceleration due to the scalar self-force, and the second term describes the coordinate acceleration of the background geodesic.

The metric and Christoffel symbols are known a priori as the particle is evolved on a fixed Schwarzschild background in Kerr-Schild coordinates. The relevant expressions can be found e.g. in \cite{Visser:2007fj}. 

\subsection{Iterative Scheme}\label{sec:Evolution/Iterative}
The particle's self-acceleration is driven by spatial and time derivatives of the regular field $\Psi^\mR$, as described in Eqs.~(\ref{eq:cov_evolution_split}) or (\ref{eq:coord_evolution}).  Inside the worldtube, the regular field is represented by a Taylor expansion, the coefficients of which are determined from continuity conditions on $\Psi^\mR$ and its time derivative on the worldtube boundary, Eqs.~\eqref{eq:continuity_condition} and (\ref{eq:continuity_condition2}). These conditions involve the puncture field $\Psi^\mP$ and its time derivative, which themselves, however, depend on the particle's self-acceleration and its derivatives; recall Eq.\ (\ref{eq:PsiP_acc}). The acceleration equation~\eqref{eq:coord_evolution} is therefore an \textit{implicit} equation for $\ddot x^i_p$. 

To deal with this problem, we construct an iterative scheme. For the ease of the reader, we first define the scheme using just the geodesic component of the puncture field, $\Psi^\mP = \Psi_{\rm geo}^\mP$ and elaborate how the acceleration terms $\Psi_{\rm acc}^\mP$ are included in the next section. The geodesic puncture field $\Psi_{\rm geo}^\mP$, as given in Eq.~\eqref{eq:PsiP_geo}, only depends on the particle's position and velocity but its time derivative $\partial_t \Psi_{\rm geo}^\mP$ depends on the particle's acceleration.

Let $\ddot{x}^i_{p(k)}$ be this acceleration during the $k$-th iteration of this scheme. From this, we compute the corresponding value for the geodesic puncture field by evaluating Eq.~\eqref{eq:PsiP_geo} and its time derivative
\begin{align}\label{eq:puncture_field_iteration}
	\Psi^\mP_{(k)} &= \Psi^\mP_{\rm geo}(x^i_p, \dot{x}^i_p), \\
	\partial_t \Psi^\mP_{(k)} &= \partial_t \Psi^\mP_{\rm geo}(x^i_p, \dot{x}^i_p, \ddot{x}^i_p = \ddot{x}^i_{p(k)}).
\end{align}
This allows us to calculate iteration $k$ for the Taylor expansions of the regular field $\Psi^\mR_{(k)}$ and its time derivative $\partial_t \Psi^\mR_{(k)}$ using the continuity condition~\eqref{eq:continuity_condition} and~\eqref{eq:continuity_condition2}
\begin{align}
	\Psi^\mR_{(k)}(t, x^i) &\overset{\Gamma}{=}\Psi^\mN(t, x^i) - \Psi^\mP_{(k)}(t, x^i), \\
	\partial_t \Psi^\mR_{(k)}(t, x^i) &\overset{\Gamma}{=} \partial_t \Psi^\mN(t, x^i) - \partial_t \Psi^\mP_{(k)}(t, x^i).
\end{align}
The regular field then yields an updated guess for the acceleration through Eq.~\eqref{eq:coord_evolution}
\begin{equation}
	\ddot{x}^i_{p(k+1)} = \ddot{x}^i(\partial_\alpha \Psi^\mR_{(k)}).
\end{equation}
The updated acceleration can be re-inserted into Eq.~\eqref{eq:puncture_field_iteration} and the iteration procedure can in principle be repeated an arbitrary number of times. The convergence of this scheme is explored in Section~\ref{sec:Results/Iterative}.

We initialize this iterative procedure with the geodesic acceleration $\ddot{x}^i_{(0)} = \ddot{x}^i_{\rm geo}$ as given by the second term in Eq.~\eqref{eq:coord_evolution}. This choice conveniently separates the first iterations by order in $\epsilon$: the values from the 0-th iteration $\ddot{x}^i_{p(0)}$, $\partial_t \Psi^\mP_{(0)}$ and $\partial_t \Psi^\mR_{(0)}$ are all computed for a geodesic orbit and are accurate up to order $\epsilon^0$. The first iteration of the acceleration $\ddot{x}^i_{p(1)}$ is then accurate up to order $\epsilon^1$, as is the resulting derivative of the puncture field $\partial_t \Psi^\mP_{(1)}$.

\subsection{Evaluation of acceleration terms}\label{sec:Evolution/Acceleration}

The acceleration terms of the puncture field $\Psi_{\rm acc}^\mP$ directly depend on the particle's acceleration $\ddot{x}^i_p$ captured by the self-acceleration $f^\alpha$ and its derivatives, see Eq.~\eqref{eq:PsiP_acc}. We now explain how this contribution, and its required derivatives, are evaluated at the $k$th iteration step, given the particle's current position and velocity as well as $\Psi^\mR_{(k-1)}(t, x^i)$ and $\partial_t \Psi^\mR_{(k-1)}(t, x^i)$.

We denote the partial derivative of a field $h(t, x^i$) evaluated at the position of the particle $x_p^i$:
\begin{equation}
	\left.\frac{\partial h}{\partial x^\alpha}\right|_{x^i_p}\!\!\!(t) = \frac{\partial h}{\partial x^\alpha} (t, x^i = x^i_p(t)).
\end{equation}
We label with a tilde fields evaluated along the path of the particle, $\tilde f(t) = f(t, x^i = x^i_p(t))$. We also introduce the total time derivative operator $d_t$ to take time derivatives of fields evaluated at the position of the particle. It acts on an arbitrary field $\tilde h$ as
\begin{equation}
	d_t \tilde h= \left.\frac{\partial h}{\partial t}\right|_{x^i_p}
	+ \dot{x}^i_p \left.\frac{\partial h}{\partial x^i}\right|_{x^i_p}.
\end{equation}
The second total time derivative is given by
\begin{align}
	d^2_t \tilde h =&
	\left.\frac{\partial^2 h}{\partial t^2}\right|_{x^i_p}
	+ 2 \dot{x}^i_p \left.\frac{\partial^2 h}{\partial t \partial x^i}\right|_{x^i_p}
	+\dot{x}^i_p \dot{x}^j_p \left.\frac{\partial^2 h}{\partial x^i \partial x^j}\right|_{x^i_p}
	+ \ddot{x}^i_p \left.\frac{\partial h}{\partial x^i}\right|_{x^i_p}.
\end{align}

For $n=1$, the acceleration terms Eq.~\eqref{eq:PsiP_acc} depend on $f^\alpha(t)$ and its first covariant derivative along the orbit, $(u^\beta \nabla_\beta f^\alpha)(t)$. We also require  $\partial_t \Psi^\mP_{\mathrm{acc}}$, which involves the time derivative of these two quantities. These expressions, in turn, require the calculation of the first and second time derivatives of the four velocity, $\dot{u}^\alpha$ and $\ddot{u}^\alpha$, as well as various partial derivatives of the regular field evaluated at the position of the particle. Here we give explicit expressions for all these necessary input quantities. 

The first two time derivatives of the self-force are given by
\begin{subequations}
	\begin{align}
	\dot f^\alpha =\frac{q}{\mu} \Biggl[ &\left(d_t (\tilde g^{\alpha \beta}) + \dot{u}^\alpha u^\beta + u^\alpha \dot{u}^\beta\right) \partial_\beta \Psi^\mR|_{x^i_p} \nonumber\\
	 &+ \left(\tilde g^{\alpha \beta} + u^\alpha u^\beta \right) d_t (\partial_\beta \Psi^\mR|_{x^i_p}) \Biggr], \\
	\ddot f^\alpha  = \frac{q}{\mu} \Biggl[& \left(d^2_t (\tilde g^{\alpha \beta}) + \ddot{u}^\alpha u^\beta + 2 \dot{u}^\alpha \dot{u}^\beta + u^\alpha \ddot{u}^\beta\right) \partial_\beta \Psi^\mR|_{x^i_p} \nonumber\\
	\nonumber &+ 2 \left(d_t (\tilde g^{\alpha \beta}) + \dot{u}^\alpha u^\beta + u^\alpha \dot{u}^\beta \right) d_t (\partial_\beta \Psi^\mR|_{x^i_p}) \\
	 &  + \left(\tilde g^{\alpha \beta} + u^\alpha u^\beta \right) d^2_t (\partial_\beta \Psi^\mR|_{x^i_p}) \Biggr].
	\end{align}
\end{subequations}
The derivative of $u^\beta \nabla_\beta f^\alpha$ is given by
\begin{align}
	\frac{d}{dt} (u^\beta \nabla_\beta f^\alpha) &= \ddot{f}^\alpha u^0 + \dot{f}^\alpha \dot{u}^0\\
	\nonumber &+ d_t\tilde \Gamma^\alpha_{\beta \gamma} u^\beta f^\gamma + \tilde \Gamma^\alpha_{\beta \gamma} \dot{u}^\beta f^\gamma + \tilde \Gamma^\alpha_{\beta \gamma} u^\beta \dot{f}^\gamma.
\end{align}
The quantities $\partial_i g^{\alpha \beta}$, $\partial_i \partial_j g^{\alpha \beta}$ and $\partial_i \Gamma^\alpha_{\beta \gamma}$, which are required for the total time derivatives of the metric and Christoffel symbols, are calculated analytically. We do not give the expressions here for brevity.

The first derivative of the four velocity is given directly by the evolution equation~\eqref{eq:cov_evolution_split}, and the second time derivative can be obtained from its derivative:
\begin{subequations}
\begin{align}
	\dot{u}^\alpha &=\frac{1}{u^0} \left(\frac{q}{\mu} \tilde g^{\alpha \beta} \partial_\beta \Psi^\mR|_{x^i_p} - \tilde \Gamma^\alpha_{\beta \gamma} u^\beta u^\gamma \right), \\
	\ddot{u}^\alpha &=\frac{1}{u^0} \bigg(\frac{q}{\mu} d_t \tilde g^{\alpha \beta} \partial_\beta \Psi^\mR|_{x^i_p} + \frac{q}{\mu} \tilde g^{\alpha \beta} d_t(\partial_\beta \Psi^\mR|_{x^i_p})  \nonumber\\
	 & \qquad\quad\ - d_t \tilde\Gamma^\alpha_{\beta \gamma} u^\beta u^\gamma - 2 \tilde \Gamma^\alpha_{\beta \gamma} \dot{u}^\beta u^\gamma - \dot{u}^0 \dot{u}^\alpha \bigg).
\end{align}
\end{subequations}
Some of the required derivatives of the regular field $\Psi^\mR$ can be obtained directly from the Taylor expansions~\eqref{eq:psi_expansions},
\begin{align}
	\partial_i \Psi^\mR|_{x^i_p} &= \Psi^\mR_i(t), \\
	\partial_t \Psi^\mR|_{x^i_p} &= (\partial_t\Psi^\mR)_0(t), \\
	\partial_t \partial_i \Psi^\mR|_{x^i_p} &= (\partial_t\Psi^\mR)_i(t).
\end{align}
Higher derivatives, however, are not obtainable directly in this manner. We make use of the fact that we can take arbitrarily high spatial derivatives of the regular field and its time derivative by taking spatial derivatives of their expansions. For expansion order $n=1$, this implies that all second and higher spatial derivatives of the regular field and its time derivative can be consistently set to zero.
This leaves the higher time derivatives $\partial^2_t \Psi^\mR|_{x^i_p}$, $\partial^3_t\Psi^R|_{x^i_p}$ and $\partial^2_t \partial_i \Psi^\mR|_{x^i_p}$ to be determined. We obtain these by taking derivatives of the vacuum scalar wave equation which the regular field satisfies. As they express the second time derivative in terms of spatial derivatives, we can consistently express all second time derivatives in terms of spatial derivatives yielding
\begin{align}\label{eq:d2t_PsiR}
	\partial^2_t \Psi^\mR  &= \frac{1}{g^{tt}}\Big(-2 g^{ti}\partial_t \partial_i \Psi^\mR + \Gamma^{t}\partial_t \Psi^\mR + \Gamma^{i}\partial_i \Psi^\mR \Big), \\
\partial^2_t \partial_i \Psi^\mR &= \frac{1}{g^{tt}} \Big(2 \partial_i g^{tj} \partial_t \partial_j \Psi^\mR 
	- \partial_i \Gamma^t \partial_t \Psi^\mR \\
	\nonumber & \qquad\quad- \Gamma^t \partial_t \partial_i \Psi^\mR - \partial_i \Gamma^j \partial_j \Psi^\mR -\partial_i g^{tt} \partial^2_t\Psi^\mR\Big).
\end{align}
The time derivative of Eq.~\eqref{eq:d2t_PsiR}, yields the final necessary term
\begin{equation}
	\partial^3_t\Psi^R = \frac{1}{g^{tt}} \left( 2 g^{ti} \partial^2_t \partial_i \Psi^\mR - \Gamma^t \partial^2_t \Psi^R - \Gamma^i \partial_t \partial_i \Psi^\mR \right).
\end{equation}
At this point, we have prescribed all quantities necessary for evaluating the acceleration term $\Psi^\mP_{\mathrm{acc}}$ and its time derivative in terms of the expansion coefficients of $\Psi^\mR$ and $\partial_t\Psi^\mR$, the current position and velocity of the particle, and background quantities. This allows for $\Psi^\mP_{\mathrm{acc}}$ to be included in the iterative scheme consistently. We construct the $k$th iteration of the puncture field from an acceleration $\ddot{x}^i_{p(k)}$ as
\begin{align}\label{eq:puncture_field_acc_iteration}
	\Psi^\mP_{(k)} =\;& \Psi^\mP_{\mathrm{geo}} \left(x^i_p, \dot{x}^i_p\right) \\
	\nonumber &+ \Psi^\mP_{\mathrm{acc}} \left(x^i_p, \dot{x}^i_p,  \ddot{x}^i_{p(k)}, \partial_\alpha \Psi^\mR_{(k-1)}|_{x^i_p} \right), \\
	\partial_t \Psi^\mP_{(k)} =\;& \partial_t \Psi^\mP_{\mathrm{geo}} \left(x^i_p, \dot{x}^i_p, \ddot{x}^i_{p(k)} \right) \\
	\nonumber &+ \partial_t \Psi^\mP_{\mathrm{acc}} \left(x^i_p, \dot{x}^i_p, \ddot{x}^i_{p(k)}, \partial_\alpha \Psi^\mR_{(k-1)}|_{x^i_p} \right).
\end{align}
Recall that the scheme is initialized with the geodesic acceleration $\ddot{x}^i_{p(0)} = \ddot{x}^i_{\rm geo}$ so the zeroth iteration of the puncture field is given by its geodesic component $\Psi^\mP_{(0)} = \Psi^\mP_{\mathrm{geo}}$ and $\partial_t \Psi^\mP_{(0)} = \partial_t \Psi^\mP_{\mathrm{geo}}$. The acceleration terms $\Psi^\mP_{\mathrm{acc}}$ only start to contribute to the particle's acceleration at the second iteration $\ddot{x}^i_{p(2)}$. At this point, $\ddot{x}^i_{p}$ includes terms of $\Order (\epsilon^2)$ and we must use in Eq.~\eqref{eq:coord_evolution} the dynamical mass $\mu(t)$ as obtained in Eq.~\eqref{eq:dynamic_mass}, rather than the rest mass $\mu_0$.

The acceleration terms can cause the simulation to grow unstable during a self-consistent evolution. We find this happens only for relatively large $\epsilon\, (\gtrsim 0.1)$ and far into the inspiral, usually when the scalar charge is near the horizon of the central black hole. The instabilities do not occur when the acceleration terms are not included in the evolution. We are unsure what the underlying cause is, but it does not affect the regions of parameter space we probe in Sec.~\ref{sec:Results}. It might be a consequence of the coupled system (the Klein-Gordon equation coupled to the particle's equation of motion) being effectively higher than second-order in time, meaning the instability could be similar in nature to the well-known problem of runaway solutions in the equation of motion of an accelerated charged particle in electromagnetism~\cite{Dirac:1938nz,Spohn:1999uf}; our iterative procedure is similar to an iterative reduction-of-order approximation in that context~\cite{Ekman:2021eqc}. If the instability is a pathology of the original system of equations~\eqref{eq: KG with source} and \eqref{eq:cov_evolution} in this way, then it can be understood as a failing of the point-particle approximation~\cite{2008PhRvE..77d6609R,Gralla:2009md}.

Spectre employs a task-based parallelism design where the worldtube and the DG elements are assigned to different cores of a computational cluster. At each iteration, the DG elements neighboring the worldtube evaluate the puncture field $\Psi^\mP_{(k)}$, integrate it over the worldtube boundary and send the result to the worldtube. It uses this data to compute the next iteration of the acceleration, $\ddot{x}^i_{p(k+1)}$, and the self-force per unit-mass $f^\alpha$ and its derivatives, which it then sends back to the neighboring elements. Each iteration therefore introduces a synchronization point between the worldtube and adjacent DG elements where computational cores are idly waiting for the results of another core. We find that in our simulations each iteration does increase runtime by $15-20$ per cent compared to the evolution with one iteration.

\subsection{Evolving the Orbit}

Given the position $x_p^i(t_s)$ and velocity $\dot{x}_p^i(t_s)$ of the particle, as well as data for the evolved fields $\Psi^\mN(t_s, x^i)$, $\Pi(t_s, x^i)$ and $\Phi_i(t_s, x^i)$ at time step $t_s$, we can evaluate time derivatives of the evolved fields using Eq.~\eqref{eq:KG_vacuum} and compute the acceleration $\ddot{x}_p^i(t_s)$ with Eq.~\eqref{eq:coord_evolution}. Both the PDEs for the fields and the ODEs of the trajectory are advanced with the same time stepper and step.

The evaluation of the evolution equations requires for the DG method to know both the position and velocity of the collocation points at the corresponding time step as discussed in Section~\ref{sec:Maps}. These are set by the new position $x_p^i(t_{s+1})$ and velocity $\dot{x}_p^i(t_{s+1})$ of the particle through the time-dependent parameters $\lambda_r(t_{s+1})$ and $\phi(t_{s+1})$ and their derivatives, by demanding that the center of the worldtube is mapped onto the new position of the worldtube through Eqs.~\eqref{eq:fix_fots}. The parameters controlling the excision sphere radii $\lambda_{\mathrm{wt}}(t_{s+1})$ and $\lambda_{\mathrm{bh}}(t_{s+1})$ are fixed by the condition~\eqref{eq:excision_fots}. These fully determine the global map from grid to inertial Kerr-Schild coordinates $x^i(t_{s+1}, x^\ibar)$ and its time derivative at the new time step $t_{s+1}$ and therefore the position and velocity of each grid point. Both the evolved variables and the orbit can now be advanced to the next time step $t_{s+2}$ and the procedure repeated. When using multi-step methods, all variables are updated each substep.

\section{Results}\label{sec:Results}
For the results presented here, we excise a sphere with initial radius $R_0 = 1.99 M$ from the center of our domain. The excision will at all times be contained within the event horizon of the Schwarzschild black hole of fixed mass $M$. The particle with charge $q$ and mass $\mu$ is initially placed at an orbital radius $r_0$  and the worldtube is centered on it. The outer boundary of the domain is placed at $r = 500M$. The left side of Fig.	~\ref{fig:grid} shows a cut through the inner part of the domain at the start of the simulation.

The DG evolution of the scalar field $\Psi^\mN$ is carried out in a manner very similar to that of Paper II. We employ a multi-step fourth-order Runge-Kutta method \cite{Owren1992} and the orbital parameters are advanced along with the evolved variables at each substep. A weak exponential filter is employed on all the evolved variables at each time step. The resolution of the DG grid is always set so that its error is subdominant to the error introduced by the worldtube. For simplicity, we always choose the charge and mass to be equal: $q = \mu_0 = \epsilon M$. This is no restriction, since only the ratio $q^2 / (\mu_0 M)$ is relevant for the evolution of the system.

At time $t = t_0 = 0$, the regular field inside the worldtube and the evolved variables are set to zero throughout the domain, $\Psi^\mR(t_0, x^i) = \Psi^\mN(t_0, x^i) = \Pi(t_0, x^i) = \Phi_i(t_0, x^i) = 0$. The simulation is then evolved up to $t_1 = 1500M$ with the worldtube orbiting on a prescribed, circular geodesic exactly as in Paper II. During this time, $\Psi^\mR$ and the evolved variables converge to a steady-state solution which acts as initial condition to the inspiral.

Starting at $t_1$, we include the acceleration due to the scalar self-force given by the first term in Eq.~\eqref{eq:coord_evolution}, as discussed in Section~\ref{sec:Evolution}. A transition function $w(t)$ is used to continuously activate this extra term, chosen as
\begin{equation}
	w(t) = 1 - \exp{ \left(- \left(\frac{t - t_1}{\sigma} \right)^4 \right) }.
\end{equation}
Here, $\sigma$ is the timescale over which the scalar self-force is turned on. A short timescale will cause the orbit to have higher residual eccentricity whereas a long time scale is computationally more expensive. Quantitative results below are presented starting at $t=t_1 + 2 \sigma$ where the self-force is fully active to more than one part in $10^7$. The smoothness of the turn-on function also avoids a jump in the puncture field caused by the addition of the acceleration terms $\Psi_{\rm acc}^\mP$.

\begin{figure}
	\includegraphics[width=0.98\columnwidth]{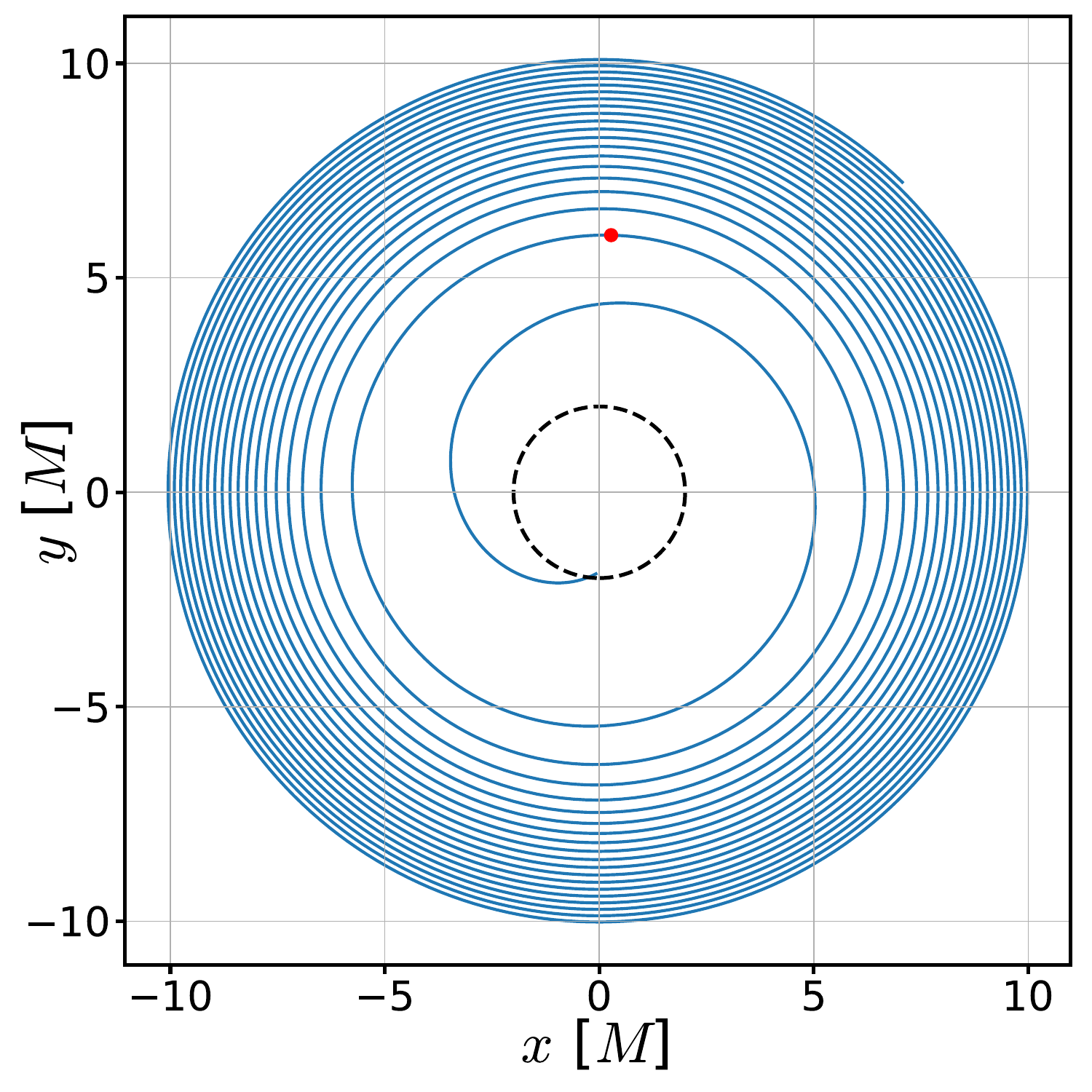}
	\caption{The orbit of a particle inspiralling into a central black hole under the influence of the scalar self-force. The orbit inspirals at a faster rate as the particle approaches the black hole and, after crossing the ISCO, plunges into the event horizon depicted by the dashed, black ring. The red dot shows the position of the particle as it crosses the ISCO at $r_p = 6M$. Here, $\epsilon = 0.08$ and $R_0 = 0.8 M$.}
	\label{fig:orbit}
\end{figure}

The back-reaction of the scalar radiation causes the charge to lose potential energy and to spiral into the central black hole on a quasi-circular orbit. Figure~\ref{fig:orbit} depicts one of these simulations. Here, the particle was placed at an initial radius $r_0=10.5 M$ and a comparatively large value $\epsilon = 0.08$ leads to a fast inspiral. The orbit is only shown starting at $t = 3500M$, at which point the self-force is fully activated. The orbital radius decreases at a faster rate as the particle gets closer to the central black hole. The red dot shows the particle's position as it crosses the ISCO at $r = 6M$. At this point, it quickly plunges into the event horizon, depicted by the dashed black line.

Once the scalar charge is contained entirely within the horizon it can no longer transfer any information to future null infinity, and there starts a vacuum ``ringdown'' evolution, in which the scalar field $\Psi^\mN$ evolves outside the BH without a source term. In practice, we choose to evolve the simulation with the scalar charge until the worldtube excision sphere is contained entirely within a radius of $r = 1.99 M$. At this point, we halt the simulation and save the values of the evolved variables $\Psi^\mN$, $\Pi$ and $\Phi_i$ on the final timeslice $t=t_{\mathrm{rd}}$. The evolution of the scalar field is continued on a new domain which has a single central excision sphere of constant radius $r =1.995 M$. This choice places the boundary within the black hole horizon so no boundary conditions are required, and outside the worldtube so data can be supplied entirely from $\Psi^\mN$. The ringdown evolution of the scalar field is then initialized at time $t_{\mathrm{rd}}$ by interpolating the evolved variables  to the new grid points. The simulation is continued on the same background spacetime for a duration of $1000 M$ at which point the scalar field has dissipated beyond the resolution of the grid. 

\begin{figure}
\includegraphics[width=\columnwidth,trim=10 5 0 0]{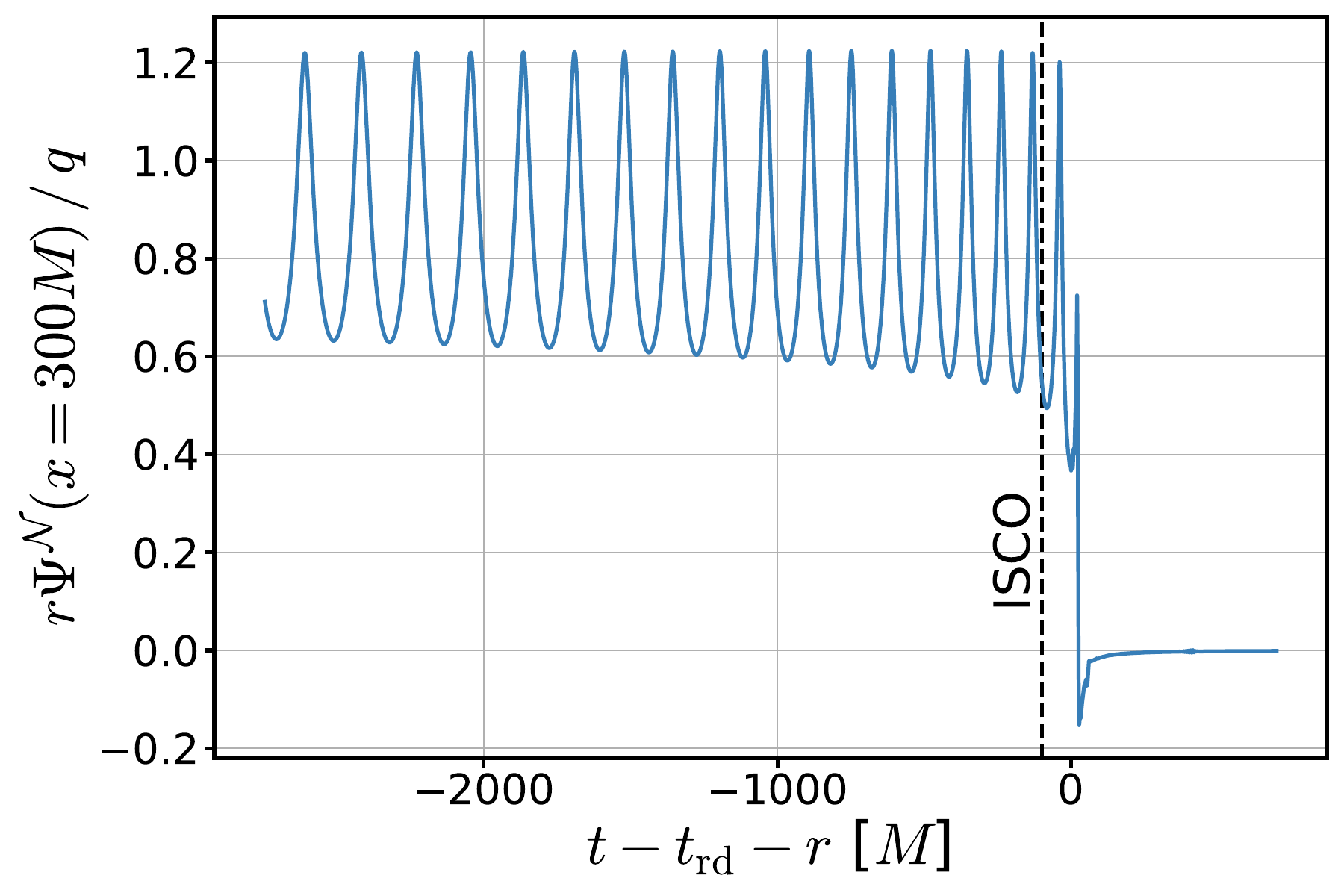}
	\caption{The value of the scalar field $r\Psi^\mN/q$ evaluated in the orbital plane at $x = 300 M$ for the same simulation as depicted in Figure~\ref{fig:orbit}. The $x$ axis shows the retarded time $t - r$, zeroed at the ringdown time $t_{\rm rd}$. The dashed vertical black line indicates the retarded time when the charge crossed the ISCO.}
	\label{fig:waveform}
\end{figure}

Figure~\ref{fig:waveform} shows the value of $r\Psi^\mN(t, x^i) / q$ in the orbital plane evaluated at $x = 300 M$. It is plotted against retarded time $t - r$ zeroed at the onset of ringdown time, corresponding to KS time $t = t_{\mathrm{rd}}$. 
The dominant frequency of the produced waveform matches the orbital frequency of the particle and gradually increases during the inspiral. The vertical dashed line shows the retarded time at which the particle crossed the ISCO. The waveform looks different to typical gravitational waveforms, as the scalar charge emits dominant monopole and dipole radiation causing its profile to oscillate around a positive value. The average amplitude of the waveform also slightly decreases during the final orbits, presumably because a significant part of the monopolar radiation is absorbed by the central black hole at this stage.

We define the phase $\phi_p(t)$ and angular velocity $\omega(t)$ of the particle as
\begin{align}
	\phi_p &= \arctan(y_p, x_p),  \\
	\omega &= \dot{\phi}_p.
\end{align}
The definition of the phase $\phi_p$ coincides with the time-dependent parameter $\phi(t)$ of the rotation map~\eqref{eq:rotation_map} because we demand that the worldtube excision sphere is tracking the particle through Eq.~\eqref{eq:fix_fots}.

In general, we will compare two simulations at the same angular velocity $\omega(t)$. As the angular velocity is strictly monotonically increasing for the quasi-circular inspirals presented here, it can be mapped to the coordinate time $t$ one-to-one. This allows us to evaluate the difference between two simulations at a common angular velocity but still plot it against the corresponding coordinate time of one of the simulations.

We also define the quantity $r_\omega = M^{1/3} \omega^{-2/3}$, which is the radius corresponding to a perfectly circular geodesic orbit with angular velocity $\omega$. During the inspiral, the value of $r_\omega$ is typically similar to the Kerr-Schild orbital radius $r_p$ of the particle. Comparing two simulations at the same $r_\omega$ is mathematically equivalent to comparing them at the same angular velocity $\omega$ but hopefully more intuitive to the reader.

In Sec.~\ref{sec:Results/Adiabatic}, we study the dependence of the self-force driven inspiral on the small parameter $\epsilon$ by fixing the initial worldtube radius $R_0$ and varying $\epsilon$ between $0.005$ and $0.08$. In the following Section~\ref{sec:Results/Radius}, we fix $\epsilon = 0.01$ and explore the convergence with worldtube radius $R_0$ by varying it between $3.2M$ and $0.2M$. In Sec.~\ref{sec:Results/AccTerms}, we repeat these simulations but do not include the acceleration terms $\Psi^\mP_{\rm acc}$ to see how this affects the evolution. Finally, we explore the convergence of the iterative scheme in Sec.~\ref{sec:Results/Iterative} by fixing both $\epsilon = 0.01$ and the initial worldtube radius $R_0 = 0.8 M$ and iterating the acceleration $\ddot{x}^i_{p(k)}$ up to $k= 2, 3, 5$ or $7$. 

\subsection{Comparison with Adiabatic Approximation}\label{sec:Results/Adiabatic}
We explore the effect of the inspiral parameter $\epsilon$ by varying it between $\epsilon = 0.005$ and $\epsilon = 0.08$ for a total of 14 values. Two simulations are run for each value of $\epsilon$, one with initial worldtube radius $R_0=0.8 M$ and one with $R_0 = 0.4M$. For simulations with $\epsilon \leq 0.01 $, we set an initial orbital radius $r_0 = 8M$. For larger $\epsilon$, the inspiral can happen so quickly that the particle would cross the ISCO before the self-force is fully turned on. To remedy this, we appropriately set larger initial orbital radii up to $10.5 M$ such that the self-force is fully active at latest when the particle reaches an orbital radius of $r_p = 7.8M$. The worldtube radius $R(t)$ is shrunk according to Eq.~\eqref{eq:shrink} with $r_0$  fixed to $8M$, even for simulations starting at larger initial separations. This leads to simulations having the same worldtube radius at the same $r_p$, independent of initial separation. We apply the iterative scheme derived in Secs.~\ref{sec:Evolution/Iterative} and~\ref{sec:Evolution/Acceleration}  and iterate each simulation until it does not affect the final results. The turn-on timescale is set to $\sigma = 1000M$ for all simulations. The puncture field includes the acceleration terms according to Eq.~\eqref{eq:puncture_field_acc_iteration}. 

\begin{figure}
	\includegraphics[width=\columnwidth,trim=5 9 0 0]{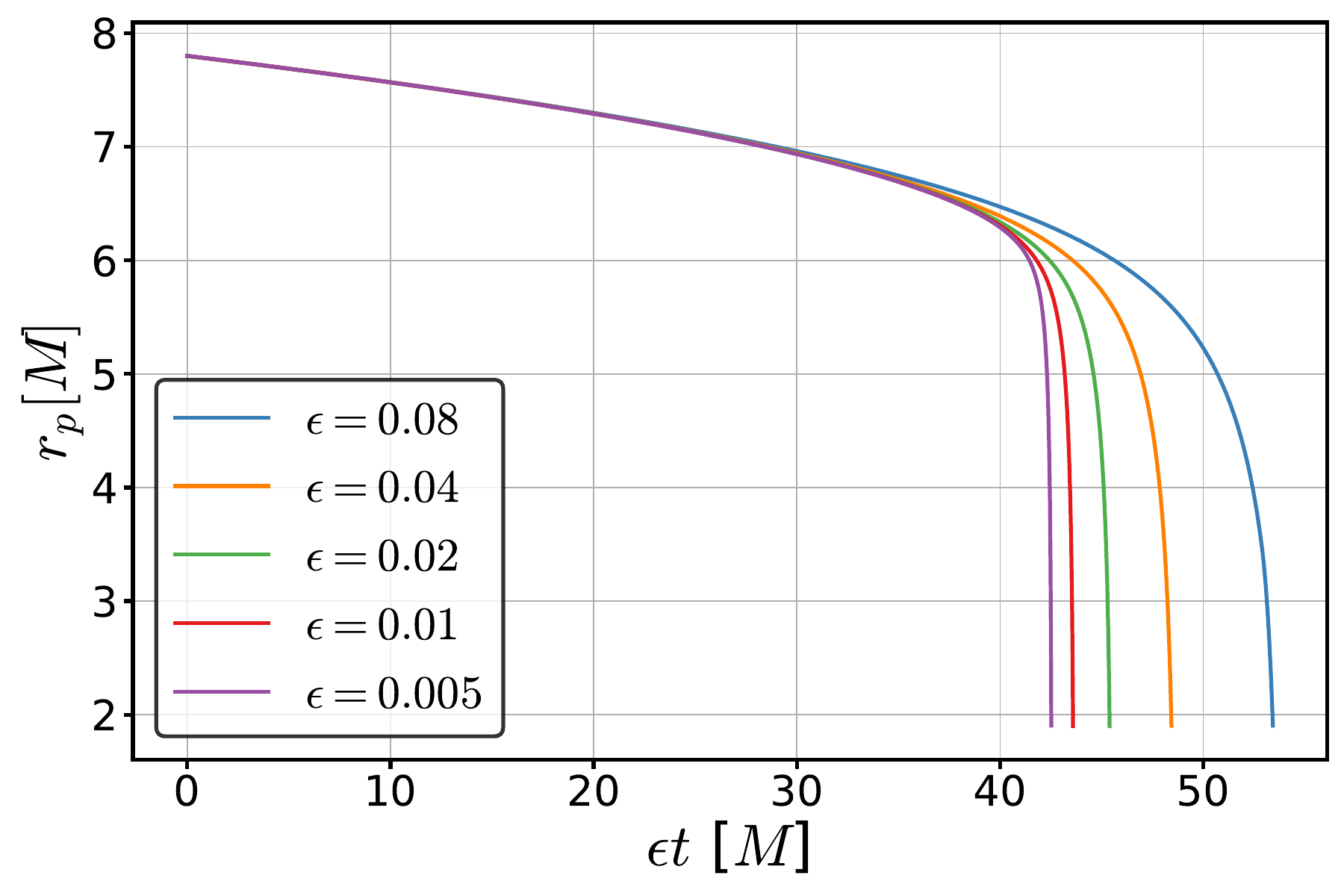}
	\caption{The orbital radius $r_p$ plotted against coordinate time $t$ multiplied by the inspiral parameter $\epsilon$ for different values of $\epsilon$. The time was set to zero for all simulations at an orbital radius of $r_p = 7.8M$, when the scalar self-force was fully active. Initially, the radii are almost identical between simulations. Near the ISCO, they start to deviate due to non-adiabatic effects. Beyond the ISCO, the particle quickly plunges into the central black hole.}
	\label{fig:inspiral_time}
\end{figure}

To understand the $\epsilon$ dependence of our results, we compare against a standard adiabatic approximation~\cite{Barack:2018yvs}, which fixes the particle on a quasi-circular orbit with $\omega = \sqrt{M/r_p^3}$ and evolves the orbital radius according to the fluxes of energy to null infinity and down the BH horizon. Concretely, we assume a solution to Eq.~\eqref{eq: KG with source} of the form
\begin{equation}
	\Psi = \sum_{\ell m} [\Psi_{\ell m}(r_p,r)+\mathcal{O}(\epsilon)]e^{-im\phi_p}Y_{\ell m}(\theta,\phi).
\end{equation}
Substituting this expansion into the Klein-Gordon equation, using $d\phi_p/dt=\omega$ and $dr_p/dt=\mathcal{O}(\epsilon)$, discarding subleading terms, and factoring out $e^{-im\phi_p}Y_{\ell m}(\theta,\phi)$ reduces the PDE to decoupled radial ODEs for the coefficients $\Psi_{\ell m}(r_p,r)$, which we solve on a grid of $r_p$ values using the Teukolsky package from the Black Hole Perturbation Toolkit~\cite{TeukolskyPackage}. At each value of $r_p$, the energy fluxes ${\cal F}_\infty$ and ${\cal F}_H$ are extracted from the solutions $\Psi_{\ell m}(r_p,r\to\infty)$ and $\Psi_{\ell m}(r_p,r\to2M)$. In terms of these fluxes, the orbital energy ${\cal E}$ changes at a rate
\begin{equation}
\frac{d\cal E}{dt} = - {\cal F} := - ({\cal F}_\infty+{\cal F}_H).  
\end{equation}
At leading order, ${\cal E}$ is related to the orbital radius by the geodesic relationship, ${\cal E}= \mu_0\frac{1-2M/r_p}{\sqrt{1-3M/r_p}}$, which allows us to express the rate of change of $r_p$ in terms of ${\cal F}$. The evolution of the orbital phase and radius are then governed by 
\begin{align}
\frac{d\phi_p}{dt} &= \omega(r_p), \label{eq:phidot 0PA}\\
\frac{dr_p}{dt} &= -\epsilon\frac{\hat{\cal F}(\hat r_p)}{d\hat{\cal E}/d\hat r_p}.\label{eq:rpdot 0PA}
\end{align}
To express the last equation in terms of $\epsilon$, we have introduced the normalized quantities $\hat{\cal E}:={\cal E}/\mu$, $\hat{\cal F}:={\cal F}/q^2$, and $\hat r_p:=r_p/M$.
\begin{figure}
	\includegraphics[width=\columnwidth,trim=10 10 0 0]{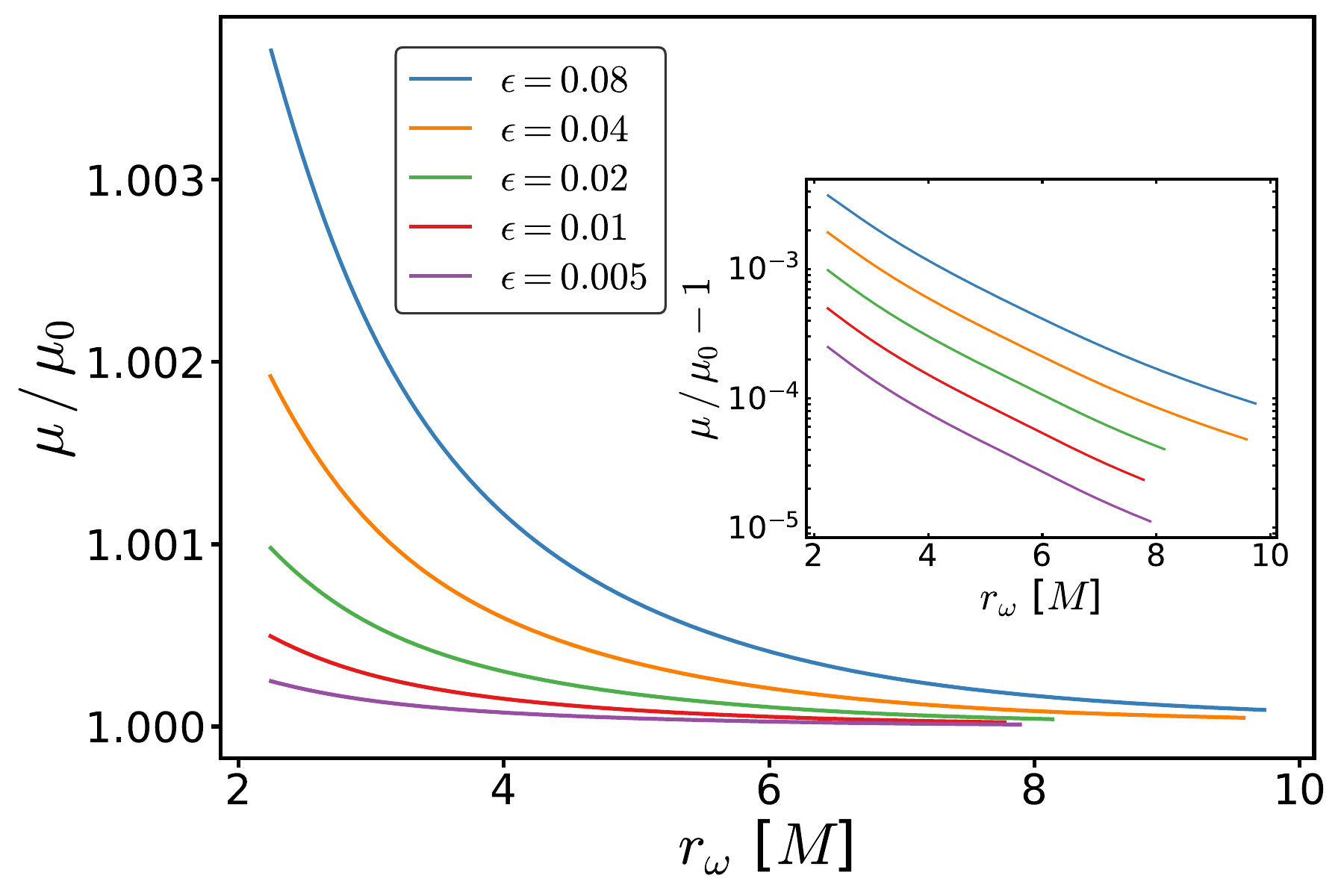}
	\caption{The evolution of the dynamic mass $\mu$ shown as a fraction of $\mu_0$ for simulations with varying $\epsilon$. The mass grows with the orbital frequency during the evolution. The inset on the right shows $\mu/ \mu_0 -1$ (plotted logarithmically), which remains proportional to $\epsilon$ throughout the simulation.}
	\label{fig:mass}
\end{figure}

The solution to Eqs.~\eqref{eq:phidot 0PA} and \eqref{eq:rpdot 0PA} can be written as
\begin{equation}
\phi_p = \frac{\phi_0(r_p)}{\epsilon}.\label{eq:phi 0PA}
\end{equation}
From Eq.~\eqref{eq:rpdot 0PA}, we can also obtain an adiabatic approximation for the dimensionless adiabaticity parameter $\dot\omega/\omega^2$,
\begin{equation}\label{eq:adiabaticity 0PA}
\frac{\dot\omega}{\omega^2} = \epsilon G_0(r_p),
\end{equation}
where $G_0(r_p)=-\frac{d\omega/dr_p}{\omega(r_p)^2}\frac{\hat{\cal F}(\hat r_p)}{d\hat{\cal E}/d\hat r_p}$. We note that unlike the self-consistent evolution we perform in our worldtube scheme, this approximation (and its extension in the next paragraph) breaks down at the ISCO: as $r_p$ approaches the ISCO, $d\hat{\cal E}/d\hat r_p$ vanishes and $G_0$ diverges.

It will also be useful to compare against the expected $\epsilon$ dependence beyond leading order in a two-timescale expansion. If the expansion is carried to higher order in $\epsilon$ following Refs.~\cite{Miller:2020bft,Miller:2023ers}, then Eqs.~\eqref{eq:phi 0PA} and \eqref{eq:adiabaticity 0PA} take the post-adiabatic form
\begin{align}
\phi_p &= \frac{\phi_0(r_\omega)}{\epsilon} + \phi_1(r_\omega) + \epsilon\phi_2(r_\omega) + \mathcal{O}(\epsilon^2),\label{eq:phase PA}\\
\frac{\dot\omega}{\omega^2} &= \epsilon G_0(r_\omega) + \epsilon^2G_1(r_\omega) + \epsilon^3G_2(r_\omega) + \mathcal{O}(\epsilon^4).\label{eq:adiabaticity PA}
\end{align}
Here it is more useful to use the invariant orbital radius $r_\omega$, but note that $\phi_0$ and $G_0$ are the same functions as in the adiabatic approximation, now simply evaluated at $r_\omega$ rather than $r_p$. 

Figure~\ref{fig:inspiral_time} shows the orbital radius $r_p$ extracted from our numerical simulations for different values of $\epsilon$ plotted against coordinate time $t$ multiplied by $\epsilon$. Here, the worldtube radius is fixed to $R_0 = 0.8M$ and we set $t$ to zero at an orbital radius of $r_p = 7.8M$ when the scalar self-force was fully active for all simulations. The rescaling of time by $\epsilon$ is motivated by Eq.~\eqref{eq:rpdot 0PA}, which shows that at adiabatic order the orbital radius is independent of $\epsilon$ when treated as a function of $\epsilon t$. Initially, our results conform to that behavior: our numerically computed orbital radii lie on top of each other, which suggests that the orbit is well described by the adiabatic approximation. The lines start to deviate near the ISCO, as non-adiabatic effects start to become significant. Once the particle passes the ISCO at $r_p = 6M$, it quickly plunges into the central black hole. The simulations shown here proceed through 69 orbits for $\epsilon =0.005$ between $r_p = 7.8M$ and $r_p = 6M$, and through 4.8 orbits for $\epsilon = 0.08$. 

Figure~\ref{fig:mass} shows the evolution of the dynamic mass $\mu$ given by Eq.~\eqref{eq:dynamic_mass}, plotted as a fraction of $\mu_0$. The mass grows as the particle inspirals and can increase by $\approx 0.1$ per cent of $\mu_0$. The inset on the right plots this fraction logarithmically as $\mu / \mu_0-1$, which can be re-written as $-\epsilon \Psi^\mR / q$ using Eq.~\eqref{eq:dynamic_mass}. It remains proportional to $\epsilon$ during the simulation as $\Psi^\mR$ is proportional to the scalar charge $q$. This again conforms to the expected behavior at adiabatic order; beyond adiabatic order, we would expect order-$\epsilon$ corrections to appear in $\Psi^\mR$ as a function of $r_\omega$, but these appear to remain small even when the particle has crossed the ISCO.

\begin{figure}
	\includegraphics[width=\columnwidth,trim=10 5 0 0]{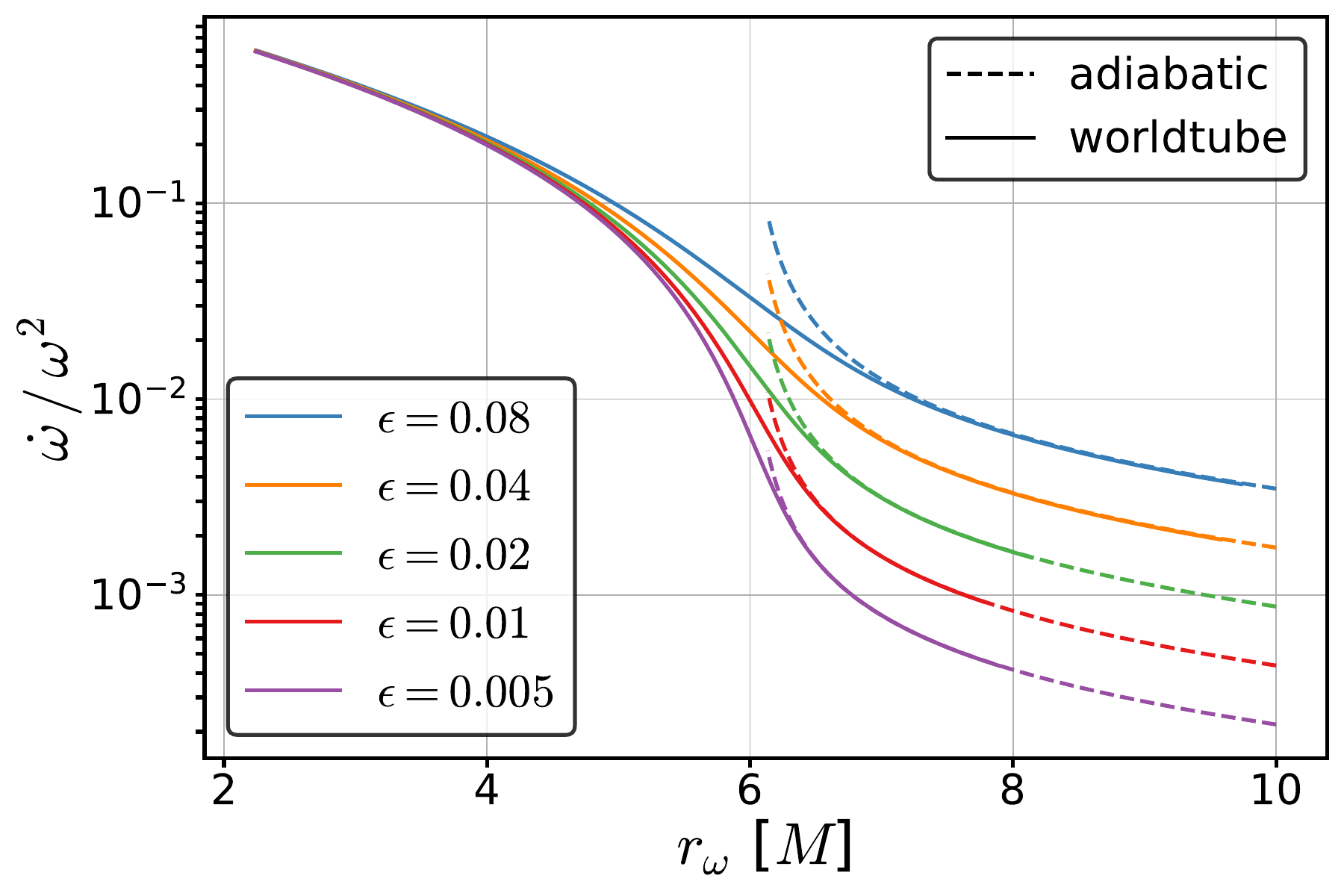}
	\caption{The value of the adiabaticity parameter $\dot{\omega} / \omega^2$ plotted for a circular inspiral for different values of $\epsilon$. The solid lines correspond to a worldtube simulation with initial worldtube radius $R_0 = 0.8M$. The dashed lines correspond to the adiabatic approximation given by the first term in Eq.~\eqref{eq:adiabaticity PA}, which breaks down at the ISCO.}
	\label{fig:adiabaticity}
\end{figure}

Figure~\ref{fig:adiabaticity} plots $\dot{\omega} / \omega^2$ against $r_\omega$. The solid lines correspond to simulations with worldtube radius $R_0 = 0.8M$. The simulations start at an initial separation between $r_0 = 8M$ and $r_0 = 10M$, depending on the value of $\epsilon$. Outside the ISCO, the particle is on a quasi-circular orbit. The geodesic angular acceleration $\dot{\omega}$ is close to zero in this regime and the adiabaticity parameter is dominated by the scalar self-force given by the first term of Eq.~\eqref{eq:coord_evolution}. The adiabaticity parameter roughly doubles for each doubling of $\epsilon$ here, as predicted by Eq.~\eqref{eq:adiabaticity PA}. For $r_p < 6M$, the scalar charge plunges into the black hole. The geodesic angular acceleration starts to dominate over the scalar self-force in this regime so the solid lines approach a common value independent of $\epsilon$.

The dashed lines in Fig.~\ref{fig:adiabaticity} show the results of the adiabatic approximation, given by the leading term in Eq.~\eqref{eq:adiabaticity PA}. We calculate these results starting at separation $r_p = 10M$ until their divergence at the ISCO. The adiabaticity parameter looks almost identical to the worldtube scheme and starts to deviate only near the ISCO.
\begin{figure}
	\includegraphics[width=\columnwidth,trim=10 5 0 0]{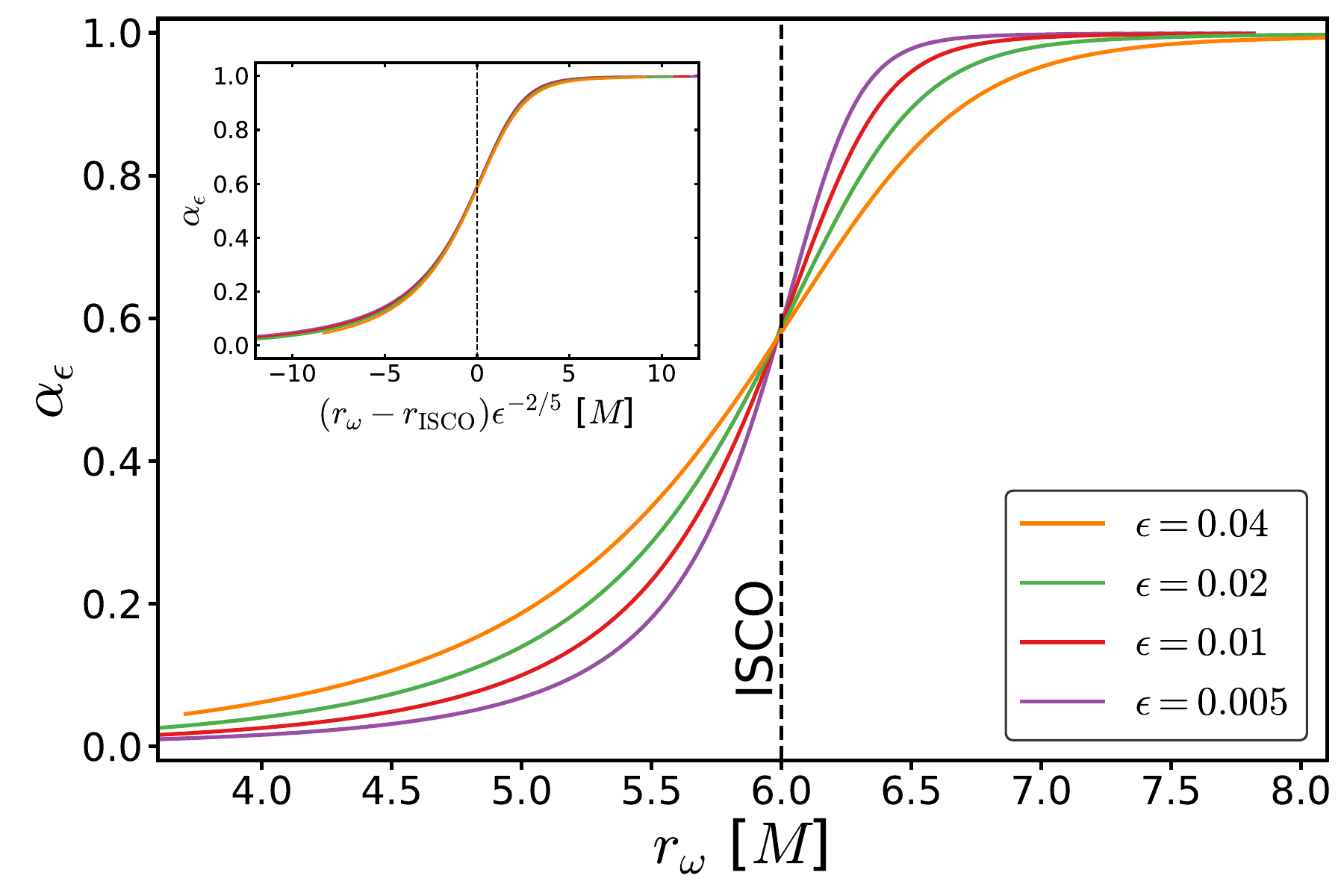}
	\caption{The convergence order $\alpha_\epsilon$ of the adiabaticity parameter $\dot{\omega}/\omega^2$ between simulations of adjacent $\epsilon$. The convergence order is close to one during the early inspiral on the right, where $\dot{\omega}/\omega^2$ is proportional to $\epsilon$ and close to zero during the final plunge on the left, where it is independent of $\epsilon$. The width of this transition grows with larger values of $\epsilon$. The inset on the left shows the same plot with the x-axis rescaled by a factor of $\epsilon^{-2/5}$ around $r_{\rm ISCO} = 6M$.  The curves now collapse, suggesting that the width of the transition regime scales as $\epsilon^{2/5}$.
  }
	\label{fig:transition_convergence}
\end{figure}

We investigate the transition regime from inspiral to plunge further and define the ``local convergence order'' of the adiabaticity parameter
\begin{equation}
	\alpha_{\epsilon, j} (r_\omega) = \frac{\log(Q_j (r_\omega)) -  \log(Q_{j-1} (r_\omega))}{\log(\epsilon_{j}) - \log(\epsilon_{j-1})},
\end{equation}
where $\epsilon_j$ are the different values of the inspiral parameter that were simulated and we have denoted with $Q_j = \dot{\omega}_j / \omega_j^2$ the corresponding values of the adiabaticity parameter.
The quantity $\alpha_\epsilon$ gives the power in $\epsilon$ with which the adiabaticity parameter changes between simulations with different $\epsilon$.  We evaluate $\alpha_\epsilon$ as a function of $r_\omega$ to investigate the different regimes inspiral, transtion-to-plunge, and plunge.
Figure~\ref{fig:transition_convergence} shows $\alpha_{\epsilon, j}$ for the same set of simulations shown in Fig.~\ref{fig:adiabaticity}. Early in the inspiral (where $r_\omega$ is significantly larger than $r_{\rm ISCO}=6M$), $\alpha_{\epsilon, j}\approx 1$, since in that regime $\dot{\omega}/\omega^2$ is proportional to $\epsilon$.  Deep inside the plunge (when $r_\omega\ll r_{\rm ISCO}$), the particle follows a plunge geodesic independent of $\epsilon$, and  $\alpha_{\epsilon, j}$ approaches zero. The transition regime in between is broader for larger values of $\epsilon$.    At the ISCO, $\alpha_{\epsilon,j}\approx 3/5$, in line with the theoretical prediction~\cite{Compere:2021zfj}; see, for example, Eq.~(25) of Ref.~\cite{Albertini:2022rfe}.

We expect the width of this interval in the transition regime to scale as $\epsilon^{2/5}$ \cite{Ori:2000zn}. To check this, we rescale the $x$-axis around the ISCO by a factor of $\epsilon^{-2/5}$ as shown in the inset of Fig.~\ref{fig:transition_convergence}.  Now the values of $\alpha_\epsilon$ coincide between the different simulations and the width of the transition region appears independent of $\epsilon$ as expected.
\begin{figure}
	\includegraphics[width=\columnwidth,trim=15 15 0 0]{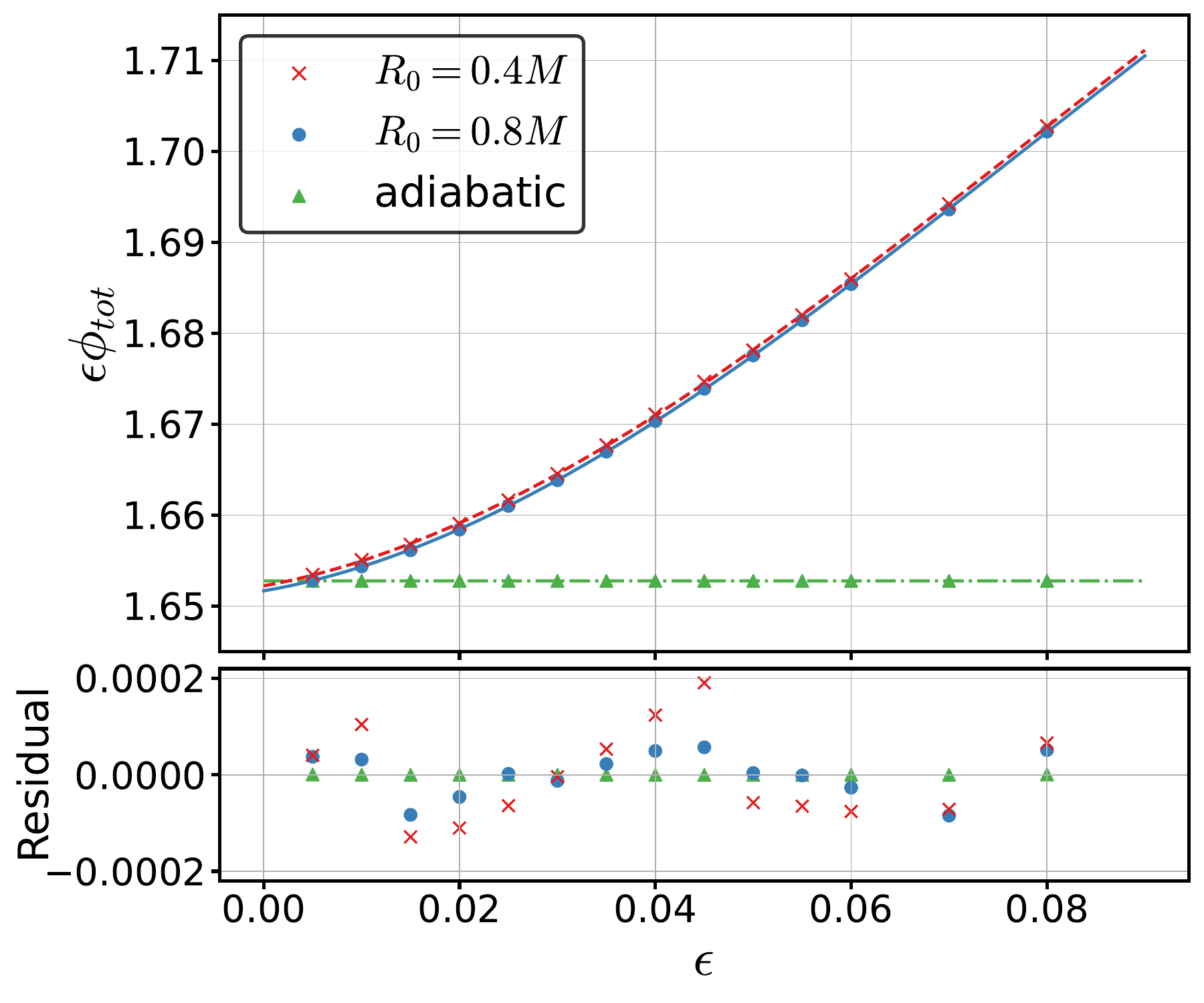}
	\caption{The total phase $\phi_{\mathrm{tot}}$ covered between the two angular velocities corresponding to  $r^{(0)}_{\omega} = 7.8 M$ and $r^{(1)}_{\omega} = 6.8 M$, multiplied by $\epsilon$. Each marker represents a separate simulation. The blue dots correspond to simulations with initial worldtube radius $R_0 = 0.8M$, the red crosses correspond to $R_0 = 0.4 M$. A cubic fit is shown for each worldtube radius as well. The green triangles correspond to the adiabatic approximation which only captures the leading order term of the scalar self-force. A linear fit is shown as well.}
	\label{fig:eps_scaling}
\end{figure}

Let us now investigate the deviations of our worldtube inspiral from the adiabatic approximation in more detail. We explore the effect of higher-order terms by considering the $\epsilon$ dependence of the orbital phase accumulated during the inspiral. We define $\phi_{\mathrm{tot}}$ as the total phase covered between the frequencies corresponding to $r^{(0)}_{\omega} = 7.8 M$ and $r^{(1)}_{\omega} = 6.8 M$. We consider an expansion of the form predicted by Eq.~\eqref{eq:phase PA},
\begin{equation}
	\epsilon \phi_{\mathrm{tot}} = a + b \epsilon + c \epsilon^2 + \ldots 
\end{equation}

Figure~\ref{fig:eps_scaling} shows $\epsilon \phi_{\mathrm{tot}}$ plotted for a range of $\epsilon$ using the adiabatic approximation (marked by green triangles), as well as for simulations with initial worldtube radius $R_0 = 0.8M$ (blue circles) and $R_0 = 0.4M$ (red crosses). Each marker corresponds to a separate simulation. Also plotted is a cubic fit for each worldtube size and a linear fit for the adiabatic approximation given by:
\begin{subequations}\label{eq:cubic_fit}
	\begin{align*}
		R_0 = 0.8M:&\enspace \epsilon \phi_{\mathrm{tot}} = 1.6516 + 0.184 \epsilon + 8.49 \epsilon^2 - 36.4 \epsilon^3,\\
		R_0 = 0.4M:&\enspace \epsilon \phi_{\mathrm{tot}} = 1.6522 + 0.195 \epsilon + 8.20 \epsilon^2 - 34.4 \epsilon^3,\\
		\text{adiabatic}:&\enspace \epsilon \phi_{\mathrm{tot}} = 1.6527 - 4.3\times 10^{-7} \epsilon .
	\end{align*}
\end{subequations}
The bottom panel displays the residuals of each fit.

The adiabatic approximation only resolves the leading order term in $\epsilon$ given by the constant coefficient $a$, whereas the worldtube simulations are sensitive to all powers of $\epsilon$. It is therefore unclear which order polynomial should be used to fit our simulations. A higher-order polynomial will always have lower residuals but will start to overfit the data at some order. A low-order fit will absorb higher-order physical effects into the low-order coefficients, skewing their values.

We choose to fit a cubic polynomial here, as the residuals start to look more or less unstructured at this point. A brief Bayesian analysis confirmed that a cubic fit has the highest evidence of all orders. The coefficients of the fit depend on the choice of the arbitrary frequencies $r^{(0)}_{\omega}$ and $r^{(1)}_{\omega}$ between which the phase is covered. We analyze the results here for $r^{(0)}_{\omega} = 7.8 M$ and $r^{(1)}_{\omega} = 6.8 M$, but our general conclusions hold for all frequency intervals examined.

The coefficient $a$ corresponds to the difference $\phi_0(r_\omega^{(1)})-\phi_0(r_\omega^{(0)})$ predicted by the adiabatic approximation~\eqref{eq:phi 0PA}. This limit is approached by the worldtube simulations at the left side of Fig.~\ref{fig:eps_scaling} as $\epsilon$ approaches zero. The cubic fits in Eq.~\eqref{eq:cubic_fit} demonstrate that the worldtube simulations extract this value with a relative error of $\sim 10^{-4}$. The smaller worldtube radius $R_0 = 0.4 M$ has a lower error indicating the expected convergence with worldtube size.

The difference in the simulations with worldtube radius $R_0 = 0.8 M$ and $R_0 = 0.4 M$ gives an estimate of the error induced by the finite size of the worldtube. The top panel of Figure~\ref{fig:eps_scaling} shows that the error remains small for the range of $\epsilon$ sampled as the points of the two simulations lie almost on top of each other. This is reflected by the small difference in the linear and post-adiabatic coefficients, which suggests that our simulations are able to resolve such higher-order effects accurately. However, the exact values of the post-adiabatic coefficients are more uncertain, as our fits yield values that depend rather strongly on the polynomial order used for the fit. An accurate extraction of such coefficients would require many simulations at small $\epsilon$ and worldtube radii, which are computationally expensive and beyond the scope of this project.

\subsection{Convergence with Worldtube Radius}\label{sec:Results/Radius}

In Paper II, we predicted the (global) converge rates $\alpha$ with worldtube radius $R$ of the numerical field $\Psi^\mN$, the regular field $\Psi^\mR$ and its derivatives $\partial_i\Psi^\mR$ which we have summarized in  Eqs.~\eqref{eq:predicted_error_psiN}--\eqref{eq:predicted_error_dpsiR}. These scaling relations were confirmed for circular geodesic orbits with radius $r_0 = 5 M$. The predicted rates are, however, valid for arbitrarily accelerated orbits and a correct generalization of the worldtube scheme should show the same behavior.

We investigate convergence with worldtube radius by running a set of simulations with $R_0$ varying between $3.2M$ and $0.2M$. As no analytical solutions or comparable codes exist to our knowledge, we choose the evolution with smallest initial worldtube radius $0.2 M$ as a reference solution and compute errors with respect to it. The initial orbital radius is set to $r_0 = 8 M$  for each simulation and we compute the second iteration of the acceleration $\ddot{x}^i_{p(2)}$ at each time step. The turn-on timescale is fixed to $\sigma = 500M$. The puncture field is computed with the acceleration terms according to Eq.~\eqref{eq:puncture_field_acc_iteration}. 

\subsubsection{Regular Field}
\begin{figure}
	\includegraphics[width=\columnwidth]{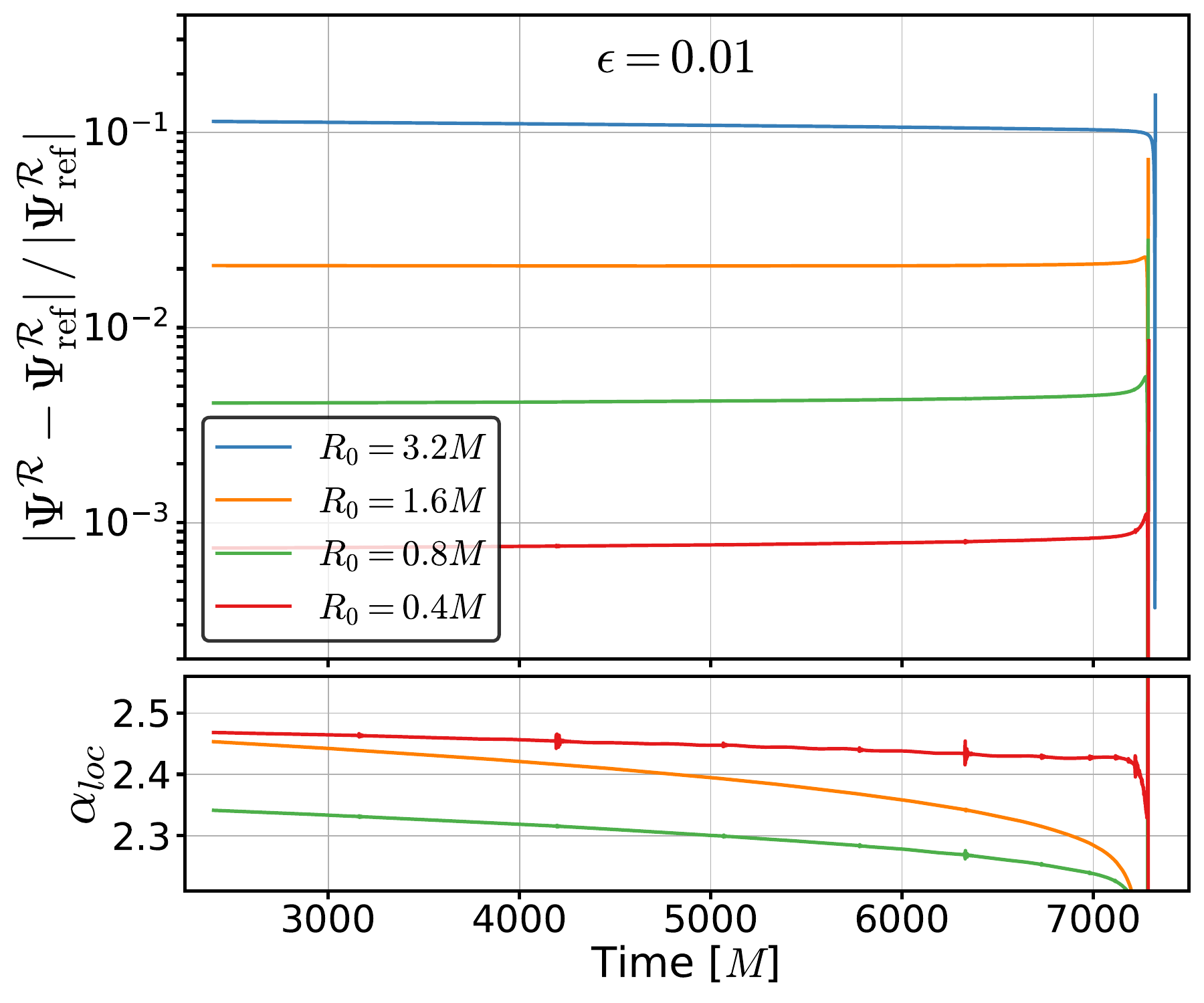}
	\caption{{\it Top panel:} The relative error of the regular field $\Psi^\mR$ at the position of the charge compared to a reference solution of small worldtube radius. The error is computed for fixed angular velocities. Each line represents a simulation with different initial worldtube radius. The error remains constant during the inspiral as the worldtube is shrunk at a rate that compensates the increasing error at smaller orbital radii.
	{\it Bottom panel:} The local convergence order between simulations of neighboring worldtube radii. It continually exceeds the expected convergence rate of $\alpha=2$.
}
	\label{fig:psi_acc}
\end{figure}
We denote the reference solution of the regular field at the position of the particle as $\Psi^\mR_{\mathrm{ref}}(\omega)$ to emphasize that we are evaluating it as a function of the particle's angular velocity. The relative error of a simulation at angular velocity $\omega$ is defined as
\begin{equation}\label{eq:psiR_error}
\varepsilon(\omega) = \frac{|\Psi^\mR(\omega) - \Psi^\mR_{\mathrm{ref}}(\omega)|}{|\Psi^\mR_{\mathrm{ref}}(\omega)|}.
\end{equation}
The top panel of Fig.~\ref{fig:psi_acc} shows this relative error $\varepsilon(\omega)$ plotted against the coordinate time corresponding to the angular velocity of the reference solution. For all simulations, the error of the regular field $\Psi^\mR$ at the charge's position remains constant until the particle is close to the event horizon. Recall that in all simulations presented here, we shrink $R(t)$ according to the power law given by Eq.~\eqref{eq:shrink} with exponent $\beta=3/2$. The constant error in the regular field confirms our hypothesis from Sec.~\ref{sec:Maps/Excision} that this choice compensates the increase in the error of the regular field $\Psi^\mR$ as the orbital radius $r_p(t)$ decreases.

The convergence rate $\alpha$ is no longer constant for the inspiralling orbits, as the simulations do not reach a steady-state solution. We introduce the local convergence order
\begin{equation} \label{eq: local convergence order}
	\alpha_{{\rm loc}, j} (\omega) = \frac{\log(\varepsilon_j (\omega)) -  \log(\varepsilon_{j-1} (\omega))}{\log(R_{0,j}) - \log(R_{0, j-1})},
\end{equation}
where $R_{0,j}$ are the different initial worldtube radii evolved, and $\varepsilon_{j}$ are the corresponding errors. The metric $\alpha_{{\rm loc}, j}$ gives a ``local" measure of $\alpha$ reached between simulations with worldtube radius $R_{0,j}$ and the next smaller worldtube radius $R_{0,j-1}$ and is therefore less prone to be influenced by zero-crossings or anomalies in the errors. The bottom panel of Fig.~\ref{fig:dphi_acc} shows $\alpha_{\rm loc}(\omega)$. The rates are continually between 2.2 and 2.5 for all worldtube radii up until the scalar charge is very close to the event horizon. This consistently exceeds the prediction $\alpha = 2$ from Eq.~\eqref{eq:predicted_error_psiR}.
\subsubsection{Angular Derivative of Regular Field}
\begin{figure}
	\includegraphics[width=\columnwidth]{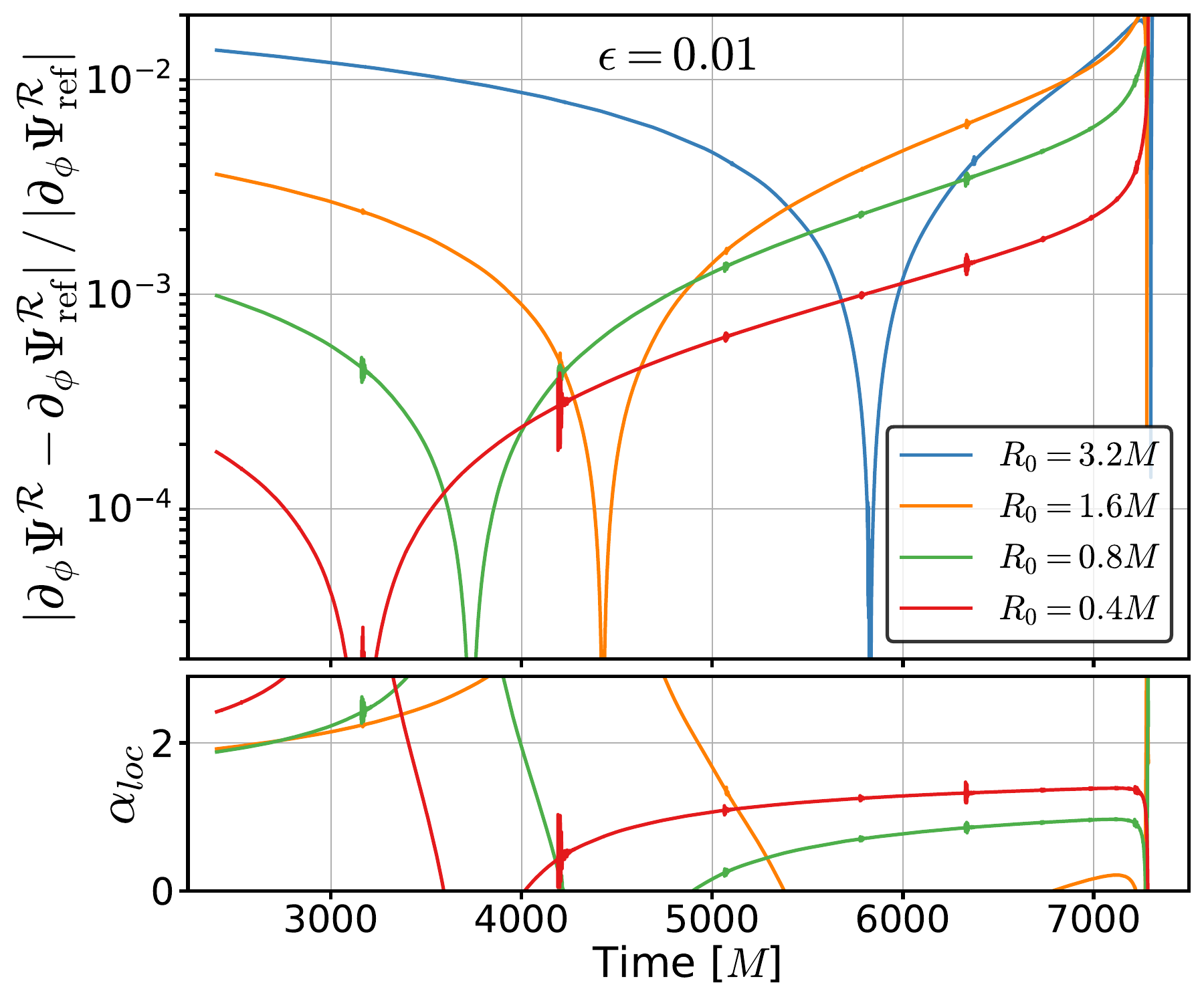}
	\caption{{\it Top panel:} The relative error of the angular derivative of the regular field $\partial_\phi \Psi^\mR$ at the position of the charge compared to a reference solution of small worldtube radius. The error is computed for fixed angular velocities.
	{\it Bottom panel:} The local convergence order between simulations of neighboring worldtube radii.
}
	\label{fig:dphi_acc}
\end{figure}
Next, we show the convergence of the error in the angular derivative of the regular field at the particle's position $\partial_\phi \Psi^\mR|_{x^i_p}$. This component corresponds to the dissipative part of the scalar self-force, which dominates the particle's inspiral rate. Its relative error against a reference simulation with initial worldtube radius $R_0 = 0.2M$ is depicted in the top panel of Fig.~\ref{fig:dphi_acc}. As before, data at the same frequency are subtracted from each other at fixed angular velocity $\omega$ but plotted against the corresponding time of the reference simulation. 

All simulations show a zero crossing in this error over the course of the inspiral, appearing later for larger worldtube radii. With the exception of this crossing, the errors are consistently increasing over the course of the evolution. This is expected as  the worldtube radius is shrunk according to the power law~\eqref{eq:shrink} with $\beta=3/2$, a choice that keeps the error in the regular field $\Psi^\mR$ constant. A value of $\beta = 3$ would be required to keep the error in the derivatives of the field constant.

The bottom panel displays the local convergence order $\alpha_{{\rm loc}}$ of the relative error in $\partial_\phi \Psi^\mR|_{x^i_p}$. When the scalar self-force is fully turned on at around $t=2500M$ at an orbital radius close to $r_p \approx 8M$, the converge order is around 2 for all simulations. As the errors goes through zero crossings, the convergence jumps but, at least for the smaller worldtube radii, appears to settle to a value $\alpha_{\mathrm{loc}} \approx 1$. We suspect that, at larger worldtube radii, higher-order terms still dominate, which causes the convergence order to be higher than predicted by Eq.~\eqref{eq:predicted_error_dpsiR}. As the orbital radius decreases, the terms stop dominating and we approach the expected convergence order. The radial and time derivative of the regular field show similar behavior, but we do not include their analysis here.

\subsubsection{Orbital Phase}
\begin{figure}
	\includegraphics[width=\columnwidth]{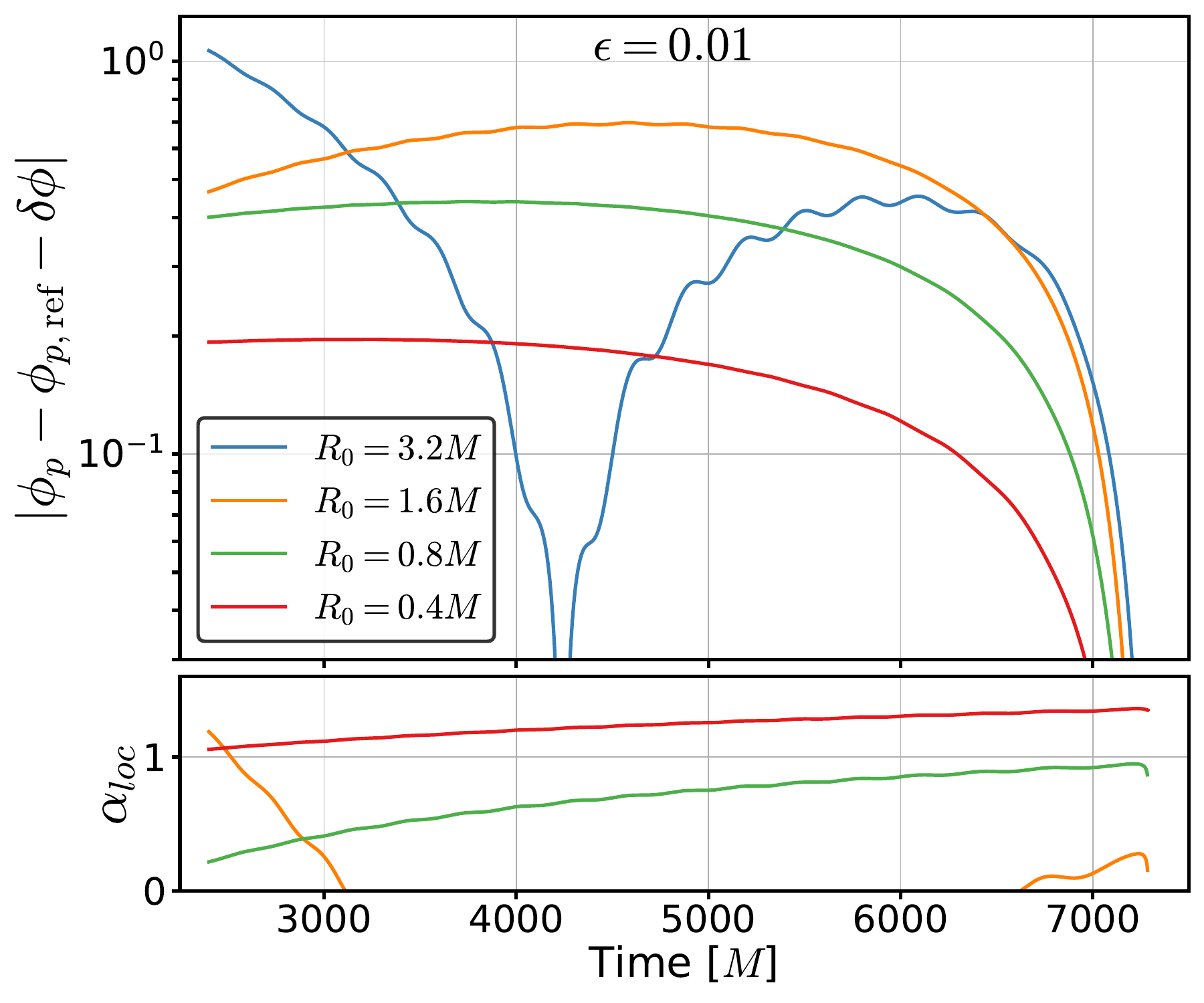}
	\caption{{\it Top panel:} The accumulated phase error of the orbit compared to a reference solution of small worldtube radius. The phase difference is set to zero at an angular velocity of $\omega_0 = 0.2 M^{-1}$ corresponding to the right end of the figure.
	{\it Bottom panel:} The local convergence order between simulations of neighboring worldtube radii. The zero crossings in the error skew the convergence orders for larger worldtube radii.
}
	\label{fig:phase_error_acc}
\end{figure}
Lastly, we consider the error in the orbital phase $\phi_p$. As the simulations are already accumulating phase differences while the self-force is being turned on, we compare phase differences at fixed angular velocity rather than time. The phase offset $\delta \phi$ and the accumulated phase error $\varepsilon(\omega)$ with respect to a reference simulation are defined as
\begin{align}\label{eq:phase_error}
	\delta \phi &= \phi_p(\omega_0) - \phi_{p,\mathrm{ref}}(\omega_0) \\
	\varepsilon(\omega) &= |\phi_p(\omega) - \phi_{p,\mathrm{ref}}(\omega) -\delta \phi|,
\end{align}
where $\omega_0$ is an arbitrary angular velocity at which the phase difference is set to zero, $\varepsilon(\omega_0) = 0$. We choose $\omega_0 = 0.2 M^{-1}$ here, which is close to the final passage through the event horizon of the particle. 

The phase differences are shown in the top panel of Fig.~\ref{fig:phase_error_acc} plotted against the coordinate time corresponding to the angular velocity of the reference simulation. The entire inspiral covered about 41 orbits while the self-force was fully turned on, during which a total phase error between $0.2$ and $1$ radians was accumulated. This corresponds to a relative error of $\sim 10^{-3}$. The phase error of the blue line with the largest initial worldtube radius of $R_0 = 3.2 M$ shows a zero crossing in the orbit as well as signs of some residual eccentricity. The orange line with initial worldtube radius $R_0 = 1.6 M$ also appears to approach a zero crossing towards the start of the simulation.

The bottom panel of Fig.~\ref{fig:phase_error_acc} shows the local convergence order $\alpha_{{\rm loc}}$. We expect that the phase error is dominated by the dissipative part of the scalar self-force driven by $\partial_\phi \Psi^\mR|_{x^i_p}$ and should therefore display the same convergence behavior of $\alpha=1$. The evolution with initial worldtube radius $R_0 = 0.4M$ shown by the red line supports this with a local convergence order slightly larger than $1$ for the entire inspiral. The zero crossings of the error in the other simulations make the analysis more difficult but the convergence order for the slightly larger initial worldtube size $R_0 = 0.8M$ appears to approach $\alpha_{{\rm loc}} \approx 1$ towards the end of the simulation.

\subsection{Effect of acceleration terms}\label{sec:Results/AccTerms}
In the previous section, we showed that the iterative scheme derived in Sec.~\ref{sec:Evolution/Iterative} attains at least the same convergence orders predicted in Eqs.~\eqref{eq:predicted_error_psiR} and~\eqref{eq:predicted_error_dpsiR} for a scalar charge inspiralling under the influence of the scalar self-force. These simulations include acceleration terms in the computation of the puncture field, and we explain our method of calculating them in Sec.~\ref{sec:Evolution/Acceleration}. Given the difficulties involved in including the acceleration terms, one might wonder whether they are indeed needed at the accuracies reached here.
\begin{figure}
	\includegraphics[width=\columnwidth]{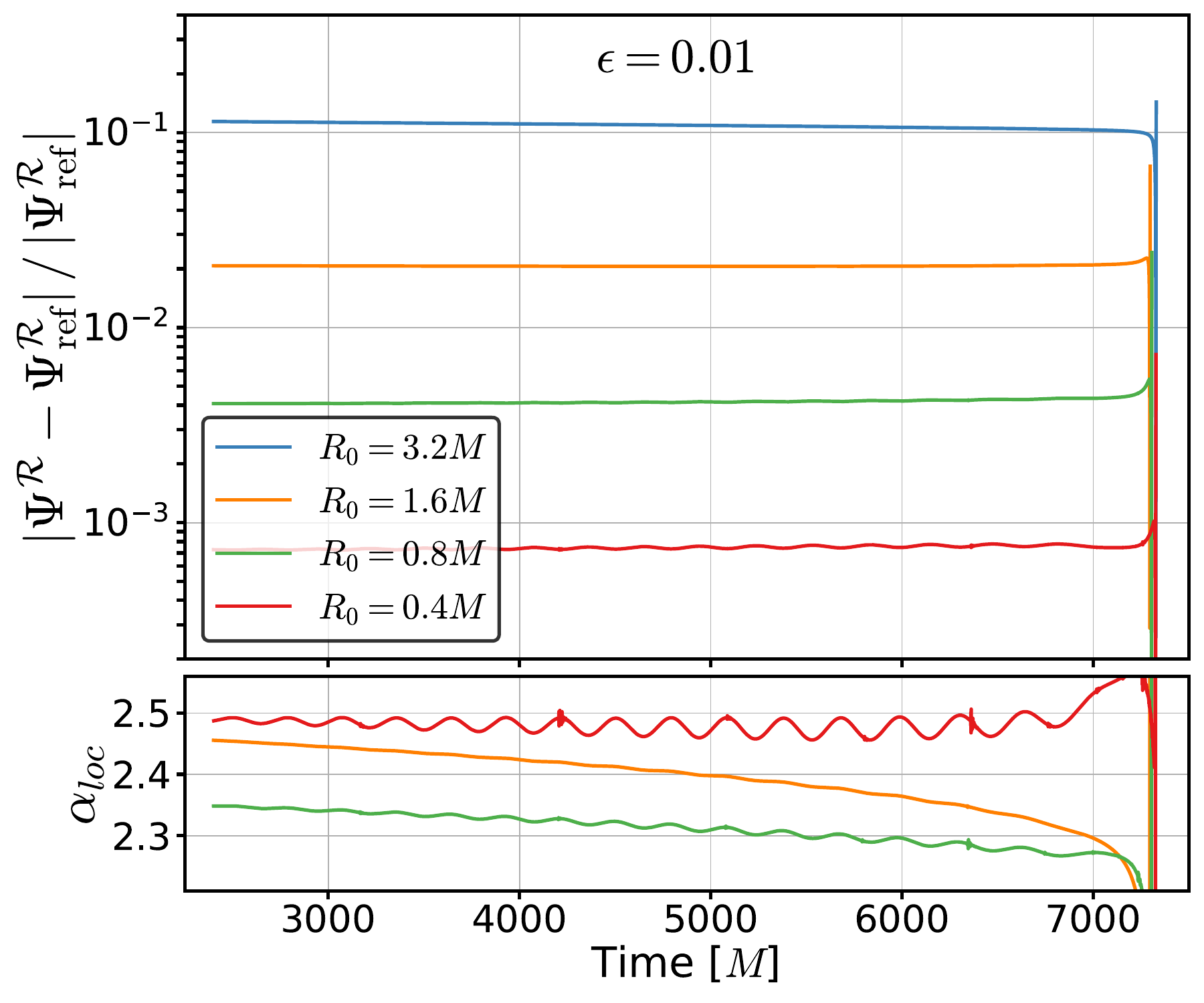}
	\caption{{\it Top panel:} The relative error of the regular field $\Psi^\mR$ at the position of the charge compared to a reference solution of small worldtube radius when not including the acceleration terms in the puncture field. 
		{\it Bottom panel:} The local convergence order between simulations of neighboring worldtube radii. This figure is very similar to Fig.~\ref{fig:psi_acc} indicating the acceleration terms do not affect the convergence rate of $\Psi^\mR$. However, they change the value to which the regular field $\Psi^\mR$ converges by about 3 per cent.}
	\label{fig:psi0}
\end{figure}
To this end, we repeat the simulations of the previous section but use Eq.~\eqref{eq:puncture_field_iteration} to evaluate the puncture field, i.e. we do not include the acceleration terms. Other than that, the simulations presented here are identical to those from the last section. The acceleration is calculated up to the second iteration $\ddot{x}^i_{p(2)}$ and the initial worldtube radius $R_0$ is varied between $3.2 M$ and $0.2 M$. The evolution with smallest initial worldtube radius $0.2 M$ (not including acceleration terms) is again used as a reference solution to compute errors with respect to it.
\subsubsection{Regular Field}
The top panel of Fig.~\ref{fig:psi0} shows the relative error of the regular field at a fixed angular velocity as defined in Eq.~\eqref{eq:psiR_error}. The error looks almost identical to the equivalent top panel of Fig.~\ref{fig:psi_acc}, which includes the acceleration terms. However, the regular field of the reference simulations $\Psi^\mR_{\rm ref}$ changes by about 3 per cent throughout the evolution if these terms are included. The regular field therefore converges to a different value.

The bottom panel shows the local convergence order $\alpha_{\rm loc}(\omega)$ between simulations with adjacent worldtube radii. These naturally also look almost identical and show convergence orders between 2.2 and 2.5, which is consistent with the prediction~\eqref{eq:predicted_error_psiR}. The only discernible effect of ignoring the acceleration terms are visible oscillations in the error and convergence rates, which suggests that the residual eccentricities between the simulations are no longer in phase. Nevertheless, we can conclude that the acceleration terms significantly change the value to which the regular field $\Psi^\mR$ converges but do not have a visible effect on the convergence rate in this regime. 

\subsubsection{Angular Derivative of Regular Field}
\begin{figure}
	\includegraphics[width=\columnwidth]{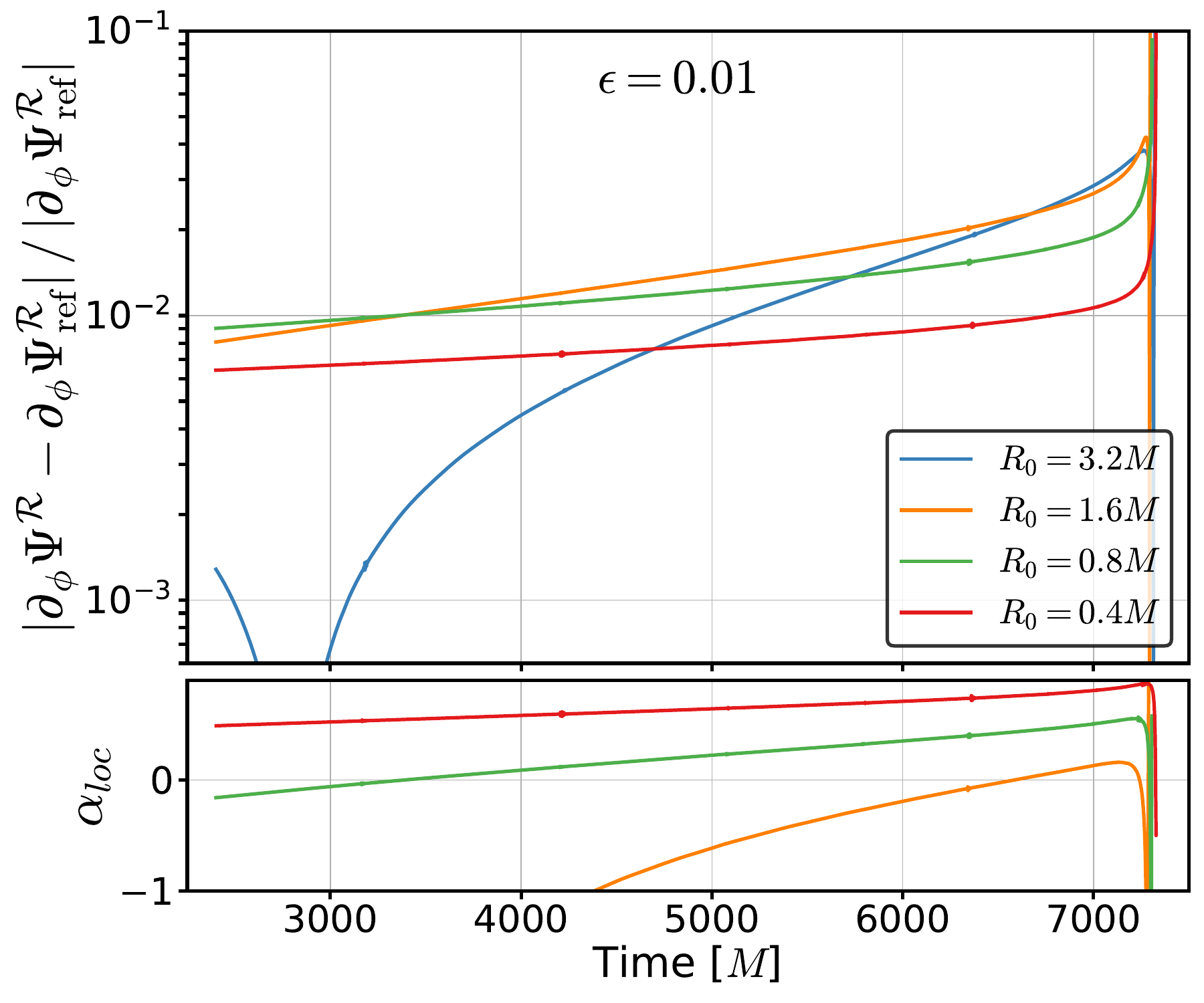}
	\caption{{\it Top panel:} The relative error of the angular derivative of the regular field $\partial_\phi \Psi^\mR$ at the position of the charge compared to a reference solution of small worldtube radius. The simulations here do not include the acceleration terms.
	{\it Bottom panel:} The local convergence order between simulations of neighboring worldtube radii. The convergence rate is below the expected value of $\alpha=1$. Figure~\ref{fig:dphi_acc} shows the same metric when the acceleration terms are included.
}
	\label{fig:dphi}
\end{figure}
Next, we explore the effect of the acceleration terms on the angular derivative of the regular field at the particle's position, $\partial_\phi \Psi^\mR|_{x^i_p}$, which is responsible for the dissipative part of the scalar self-force. Its error is as usual defined with respect to the reference solution $\partial_\phi \Psi^\mR_{\mathrm{ref}}$ as in Eq.~\eqref{eq:psiR_error}.

The relative error with respect to coordinate time is plotted in the top panel of Fig.~\ref{fig:dphi}. It is 1-2 orders of magnitude larger compared to the corresponding top panel of Fig.~\ref{fig:dphi_acc}, where the acceleration terms were included. The only exception is the largest worldtube radius, $R_0 = 3.2M$, which shows a zero crossing at the start of the simulation. For the other evolutions, decreasing the worldtube radius appears to slightly decrease the relative error in $\partial_\phi \Psi^\mR|_{x^i_p}$. The local convergence rate $\alpha_{{\rm loc}}$ depicted in the bottom panel reveals that convergence is consistently lower than the predicted rate of $\alpha = 1$.

The acceleration terms therefore appear to be essential for correctly computing the angular derivative of the regular field $\partial_\phi \Psi^\mR|_{x^i_p}$. As this component drives the inspiral of the particle, we expect that the particle's orbit to be also significantly affected.

This behavior roughly conforms with our theoretical expectation. Omitting the acceleration terms amounts to neglecting a term of order $\epsilon R^0$ in the puncture and $\epsilon R^{-1}$ in the derivative of the puncture, inducing errors of those orders in $\Psi^\mR$ and $\partial_\alpha\Psi^\mR$. Therefore, when $R\to 0$, $\Psi^\mR$ will converge but have a finite error of order $\epsilon$, while $\partial_\alpha\Psi^\mR$ will actually diverge as $R^{-1}$; the fact that we find similar convergence for $\Psi^\mR$ and slow convergence for $\partial_\alpha\Psi^\mR$ (rather than divergence) is likely due to the small value of $\epsilon$ suppressing the effect.

\subsubsection{Orbital Phase}
\begin{figure}
	\includegraphics[width=\columnwidth]{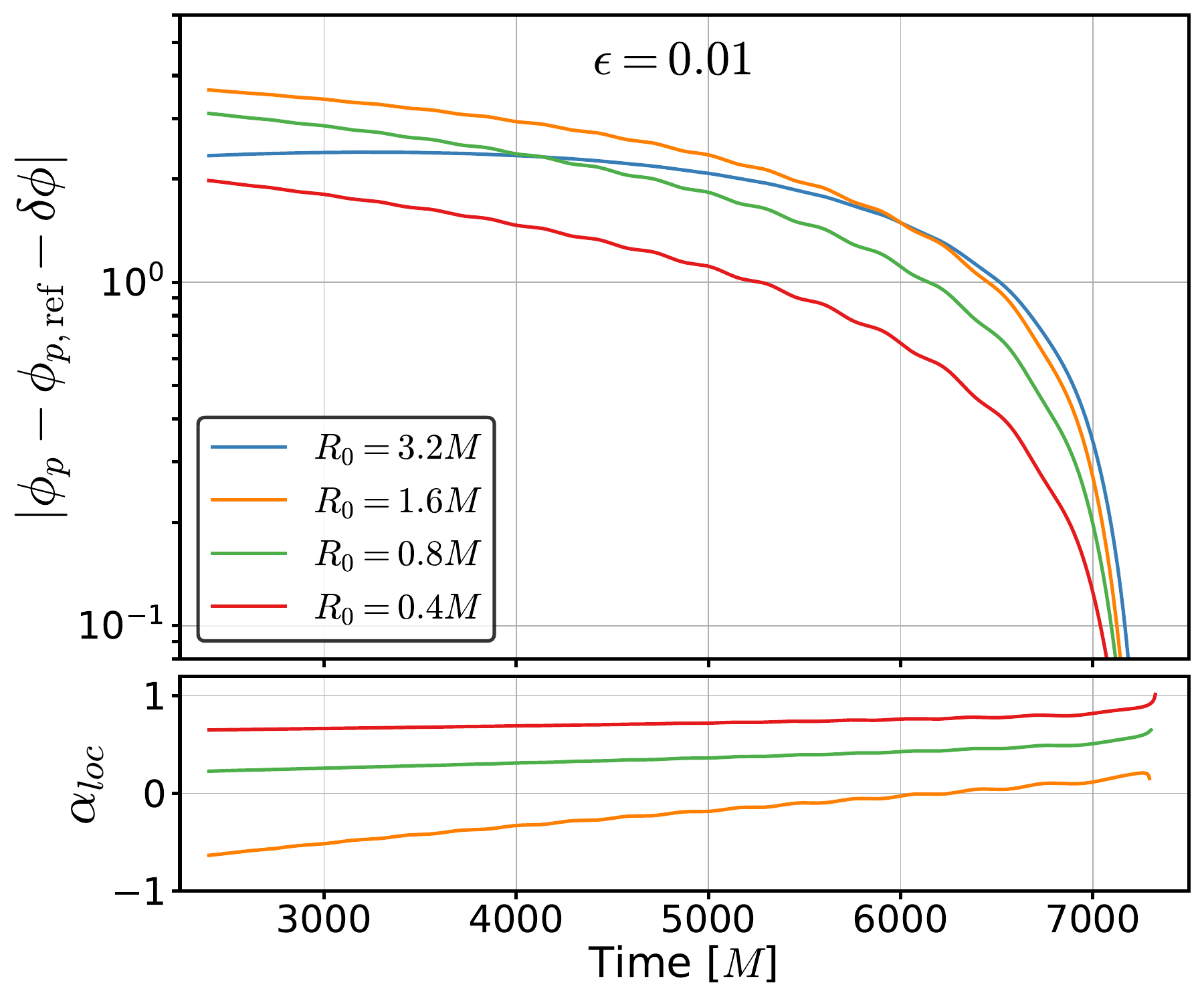}
	\caption{{\it Top panel:} The accumulated phase error of the orbit compared to a reference solution of small worldtube radius when the acceleration terms are not included. The phase difference is set to zero at an angular velocity of $\omega_0 = 0.2 M^{-1}$ which, roughly corresponding to $t \approx 7100M$.
	{\it Bottom panel:} The local convergence order between simulations of neighboring worldtube radii. Figure~\ref{fig:phase_error_acc} shows results for the same runs with acceleration terms included which have lower errors and higher convergence rates.}
	\label{fig:phase_error}
\end{figure}
Finally, we explore how the acceleration terms affect the particle's orbital phase $\phi(t)$. We measure the effect with the accumulated phase error $\varepsilon$ defined in Eq.~\eqref{eq:phase_error} which zeros the phase difference at an angular velocity $\omega_0 = 0.2M^{-1}$. The top panel of Fig.~\ref{fig:phase_error} shows $\varepsilon$ plotted against coordinate time corresponding to the angular velocity of the reference simulation. When compared to the corresponding top panel of Fig.~\ref{fig:phase_error_acc}, the accumulated phase difference is about an order of magnitude higher when the acceleration terms are omitted. While a lower worldtube radius in general still reduces the total phase difference, a comparison of the bottom panels of Fig.~\ref{fig:phase_error_acc} and~\ref{fig:phase_error} reveals that the local convergence order is consistently about half an order higher when acceleration terms are included.

\subsection{Convergence of the Iterative Scheme}\label{sec:Results/Iterative}
In Sec.~\ref{sec:Evolution/Iterative} we presented an iterative scheme that addresses the implicit form of the particle's equation of motion~\eqref{eq:coord_evolution} under the influence of the scalar self-force. We check the convergence with iterations by running a set of simulations with $k=$ 2, 3, 5 and 7 iterations of the acceleration $\ddot{x}^i_{p(k)}$, and then use the simulation with 7 iterations as a reference solution to estimate errors. We fix the initial worldtube radius to $R_0 = 0.8M$ and the inspiral parameter to $\epsilon = 0.01$. The acceleration terms are included and the turn-on timescale is set to $\sigma = 1000M$ in these runs.

\begin{figure}
	\includegraphics[width=\columnwidth]{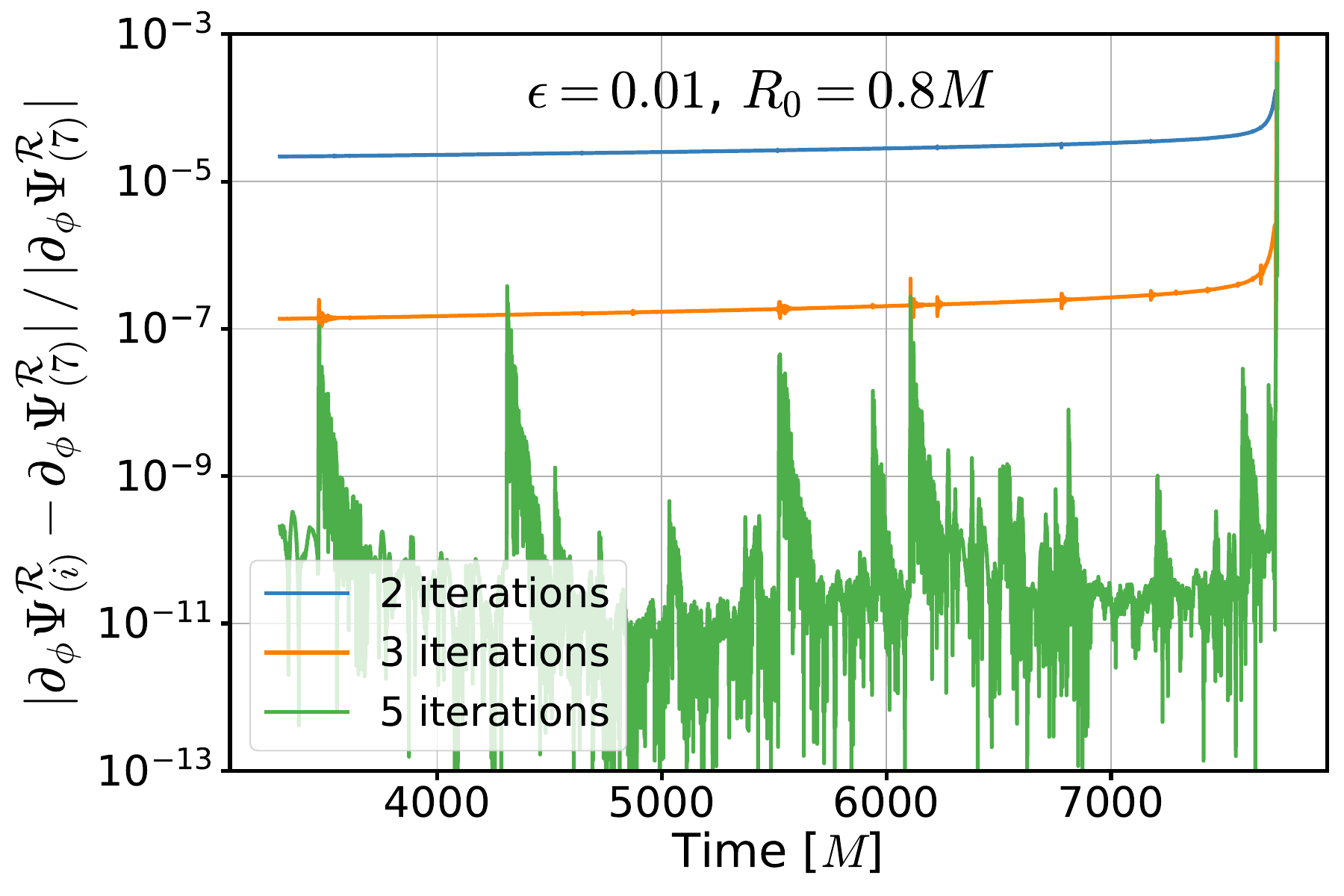}
	\caption{The relative error in the angular derivative of the regular field $\partial_\phi\Psi^\mR$ when using different number of iterations compared to using seven iterations. The error decreases by a factor of $\epsilon = 0.01$ with each iteration, as expected. The green curve is noisy due to the finite resolution of the DG evolution.}
	\label{fig:iterations}
\end{figure}

Figure~\ref{fig:iterations} shows the relative error in the angular derivative of the regular field $\partial_\phi\Psi^\mR$ at the position of the particle for 2, 3 and 5 iterations, respectively. The error is computed analogously to Eq.~\eqref{eq:psiR_error}, which compares the value at the same orbital angular velocity $\omega$ of the orbit against the reference value $\partial_\phi\Psi^\mR_{(7)}$. The simulation with 2 iterations shows a constant relative error of $\sim 10^{-5}$ until the particle is very close to the horizon. This justifies our choice of using two iterations when analyzing the convergence with worldtube radius $R_0$ in section~\ref{sec:Results/Radius} because the worldtube always induces an error at least an order of magnitude larger. 

We expect that each additional iteration adds a correction that is a factor of $\epsilon$ smaller than the previous one. This is demonstrated by the orange line, which used 3 iterations and shows an error two order of magnitudes smaller $\sim 10^{-7}$. When using 5 iterations, the additional corrections get so small that the finite resolution of the DG grid causes the error to be fairly noisy. However, a majority of the simulation still shows an error of $\sim 10^{-11}$, which is four orders of magnitude lower than with 3 iterations, as expected. For larger $\epsilon$, convergence with the iterations $k$ is slower so that several iterations were used in Sec.~\ref{sec:Results/Adiabatic} where $\epsilon$ was set as high as $0.08$.
 
\section{Conclusions and Outlook} \label{sec:Conclusions}
In this work, we continue to explore a new approach to simulating intermediate mass-ratio binary black holes in numerical relativity. The method works by excising a worldtube much larger than the smaller object from an evolution domain and employing a perturbative solution inside this region. The perturbative solution is calibrated from the evolution outside the excision sphere and in turn provides boundary conditions to it. 

In Paper II, we implemented this scheme for a scalar charge on a circular geodesic orbit using SpECTRE, a numerical relativity code that employs a DG method to evolve the Klein-Gordon equation in 3+1 dimensions. Here we extend the scheme to include the effect of radiative back-reaction on the charge, known as the scalar self-force.

We construct series of time-dependent maps that allow the worldtube to track the particle's motion on generic equatorial orbits along with the rest of the grid. Then, we derive a puncture field that is valid for generically accelerated orbits.
Finally, we show that the particle's acceleration under the scalar self-force is given in implicit form and construct an iterative scheme to address this issue.

The scheme is tested with a set of quasi-circular inspirals for different values of the inspiral parameter $\epsilon$, the worldtube radius $R$ and number of iterations $k$ used in solving the implicit equation for the self-force. We compare the results to an adiabatic approximation and show that we not only resolve effects at leading order in $\epsilon$ but also get important contributions from higher orders. We demonstrate that the regular field at the position of the particle and its derivatives converge with the worldtube radius $R$ at the theoretically predicted rates. At last, we show that the iterative scheme converges rapidly.

In this work, we have restricted ourselves to expansion order $n=1$ in coordinate distance and have shown that the resulting simulations can be run with high accuracy within a day. The inclusion of second order $n=2$, as implemented for circular orbits in Paper II, would greatly speed up simulations as a much larger worldtube radius can be used to achieve the same accuracy. Our previous work also indicates that the next order would increase the accuracy of the scheme by up to two orders of magnitude at the same worldtube radius. An implementation would require the derivation of the puncture field at the next order as well as adjusting the iterative scheme to include these higher-order terms. Both additions should be straightforward if tedious, and we leave them to future work.

While only quasi-circular orbits were presented in this work, our method is applicable for generic bound orbits. In future work, we would like to examine the effects of the scalar self-force on eccentric orbits during a self-consistent evolution. 

Other avenues for future work include the extraction of multipolar energy-momentum fluxes in scalar-field radiation to infinity and down the event horizon, which would allow us to check flux balance laws.  We currently find that the finite size of our Cauchy domain limits the accuracy at which these quantities can be extracted. This difficulty could be mitigated through a procedure of Cauchy-Characteristic extraction~\cite{Moxon:2021gbv} or Cauchy-characteristic matching~\cite{Ma:2023qjn}, in order to propagate the scalar field to null infinity.

\begin{acknowledgments}
The authors thank Peter Diener, Lorenzo Speri and Raj Patil for helpful dicussions. AP acknowledges the support of a Royal Society University Research Fellowship and the ERC Consolidator/UKRI Frontier Research Grant GWModels (selected by the ERC and funded by UKRI [grant number EP/Y008251/1]). This work makes use of the Black Hole Perturbation Toolkit. Computations were performed on the Urania HPC system at the Max Planck Computing and Data Facility.

SpECTRE uses \texttt{Charm++}/\texttt{Converse}~\cite{laxmikant_kale_2021_5597907,kale1996charm++}, which was developed by the Parallel Programming Laboratory in the Department of Computer Science at the University of Illinois at Urbana-Champaign. SpECTRE uses \texttt{Blaze}~\cite{Blaze1,Blaze2}, \texttt{HDF5}~\cite{hdf5}, the GNU Scientific Library (\texttt{GSL})~\cite{gsl}, \texttt{yaml-cpp}~\cite{yamlcpp}, \texttt{pybind11}~\cite{pybind11}, \texttt{libsharp}~\cite{libsharp}, and \texttt{LIBXSMM}~\cite{libxsmm}. The figures in this article were produced with \texttt{matplotlib}~\cite{Hunter:2007, thomas_a_caswell_2020_3948793}, \texttt{NumPy}~\cite{harris2020array}, and \texttt{ParaView}~\cite{paraview,paraview2}.

\end{acknowledgments}
		
\bibliography{references}

\end{document}